%% file: paper_0_main.tex
\documentclass[11pt]{article}	
\usepackage{amsmath,amsthm,amsfonts,amscd, amsmath}  
\usepackage{amssymb,amsfonts,amsmath,amscd,amsthm}	
\usepackage{graphicx}
\usepackage[numbers]{natbib}
\usepackage{setspace}
\usepackage[caption=false]{subfig}    
\usepackage{verbatim}      	
\usepackage{makeidx}       	
\usepackage{color}
\usepackage{enumerate}
\usepackage{authblk}
\usepackage{epsfig}         	
\usepackage{url}		
\usepackage{multirow}

\DeclareMathOperator*{\argmin}{arg\,min}
\graphicspath{{../figs/}} 


\theoremstyle{definition}

\theoremstyle{remark}

\newcommand{\latexe}{{\LaTeX\kern.125em2%
                      \lower.5ex\hbox{$\varepsilon$}}}

\chardef\bslash=`\\	

\makeatletter		
\def\square{\RIfM@\bgroup\else$\bgroup\aftergroup$\fi
  \vcenter{\hrule\hbox{\vrule\@height.6em\kern.6em\vrule}%
                                              \hrule}\egroup}
                                                                                            
\makeatother		
\makeindex    

\usepackage{fancyhdr}
\newcommand{\blue}[1]{#1}

\begin{document}
\title{Moment Preserving Constrained Resampling with Applications to Particle-in-Cell Methods}

\author[1]{D. Faghihi}
\author[3]{V. Carey}
\author[2]{C. Michoski}
\author[4]{R. Hager}
\author[2]{S. Janhunen}
\author[4]{C. S. Chang}
\author[2]{R. D. Moser}
\affil[1]{Department of Mechanical and Aerospace Engineering, University at Buffalo}
\affil[2]{Oden Institute for Computational Engineering and Sciences, University of Texas
  at Austin}
\affil[3]{Department of Mathematics and Statistics, University of Colorado Denver}
\affil[4]{Princeton Plasma Physics Laboratory, Princeton NJ}
\date{\today}
\maketitle


\begin{abstract}

The Moment Preserving Constrained Resampling (MPCR) algorithm for
particle resampling is introduced and applied to particle-in-cell
(PIC) methods to increase simulation accuracy, reduce compute cost,
and/or avoid numerical instabilities.  The general algorithm
partitions the system space into smaller subsets and resamples the
distribution within each subset. Further, the algorithm is designed to
conserve any number of particle and grid moments with a high degree
of accuracy (i.e. machine accuracy).
The effectiveness of MPCR is demonstrated with several numerical
tests, including a use-case study in gyrokinetic fusion plasma
simulations. {The computational cost of MPCR
is negligible compared to the cost of particle evolution in PIC
methods, and} the tests demonstrate that periodic particle resampling yields a significant
improvement in the accuracy and stability of the results.

\end{abstract}



\section{Introduction}\label{sec:intro}

\input{paper_1_intro}

%
%
\section{Particle Resampling Strategies and Implementation}\label{sec:method}
\input{paper_3_method}

%
\section{Numerical Examples}\label{sec:results}
\input{paper_4_results}
%
\section{Summary and Conclusions}\label{sec:conclusions}
\input{paper_5_conclusions}
%

\section*{Acknowledgments}
The work reported here was supported by the Department of Energy under
grants DE-SC0008454 and DE-AC02-09CH11466.  The authors acknowledge
the Texas Advanced Computing Center (TACC, http://www.tacc.utexas.edu)
at The University of Texas at Austin as well as the National Energy Research 
Scientific Computing Center (NERSC, http://www.nersc.gov) and Oak Ridge 
Leadership Computing Facility (OLCF, https://www.olcf.ornl.gov) supported by 
U.S. Department of Energy Office of Science for providing HPC resources that 
have contributed to the research results reported here.

\clearpage      
\bibliographystyle{unsrt}  
\bibliography{refs}
\index{Bibliography@\emph{Bibliography}}%


\end{document}

%% file: paper_1_intro.tex
Particle-in-cell (PIC) methods are a class of numerical techniques
for solving partial differential equations using a mixed Lagrangian
(particle) and Eulerian (grid cell) representation. Here we consider
PIC methods applied to systems described by the evolution of the
characteristics of a distribution function $f$. An obvious example is
Boltzmann's equation for the distribution of atomic or molecular
particles in physical and velocity space for a thermodynamic
system. Application of PIC methods is particularly natural for the
solution of Boltzmann's equation in plasmas because the mean
electromagnetic interaction of the charged atomic particles is easily
described through the Eulerian representation of electromagnetic
fields on a spatial grid. For this reason, PIC methods have been
widely used in plasma
simulations \cite{birdsall2004plasma,hockney1988,hahm1988}, both to
solve the Boltzmann equations (including atomic particle collisions)
or the Vlasov equations (neglecting atomic particle collisions). It is
this application of PIC methods to plasma simulation that motivated
the current work, but the resampling algorithms are applicable to any
{\color{red} sampled} representation of a distribution function.

In the PIC methods considered here, the marker particles that
represent the distribution function $f$ do so in a Monte Carlo sense;
that is, the marker particles are considered a random sample drawn
from the distribution $f$. In this way, expected values and
conditional expectations defined from $f$ are simply computed as
averages or conditional averages over the marker particles. While
these ``Monte Carlo PIC'' methods are simple and easy to implement,
they 
suffer from sampling error, which reduces
slowly (like $N^{-1/2}$) as the number of marker particles $N$
increases. More complex, higher-order PIC methods \cite{JACOBS200696,doi:10.1137/16M105962X}  have
also been developed. They converge more rapidly through the use of a
deterministic Lagrangian representation of $f$, but they also require
frequent remapping of the marker particles representing $f$. Here, we
pursue Monte Carlo PIC (just PIC from now on) for its simplicity, and
consider resampling algorithms to ensure that it is as accurate as
possible for a given number of particles.

A particle-in-cell method is useful because the Lagrangian particle
motion is a natural representation of the evolution of the
distribution function $f$, given the field quantities (e.g. the
electromagnetic fields in plasmas) that determine the particle
motion, while the fields are determined from $f$ (e.g. the charge and
current densities in plasmas). Thus, one of the defining features of a
PIC method is that information carried by the particles must be
projected onto the grid, where it is needed for the Eulerian solution
of the field quantities using finite
difference \cite{westermann1994}, finite volume \cite{Hermeline2008},
and/or finite element \cite{Assous1993} discretizations. Another
defining feature is that the field quantities represented on the grid
must be interpolated to the particle locations to determine the
particle evolution. A third feature of the PIC methods discussed here
is that since the particles are samples of the distribution function
$f$, they are not the same as physical particles; rather they are often
called marker particles, macro-particles, or sometimes superparticles.
For example, the marker particles are not atomic particles even when
$f$ is an atomic particle distribution; indeed, there are generally
many orders of magnitude fewer marker particles than atomic particles in
a system. These PIC features, particularly the projection of particle
information onto the grid, drive some of the requirements for the
resampling algorithms described here.

In 
Monte Carlo sampling of a distribution function $f$, one is
often concerned that low-probability (small-$f$) regions of the
phase space distribution may not be well represented, because there
will be very few particles in this region. For example, in plasma
simulations, an accurate representation may be needed of the physics
in low plasma density regions of space, or high kinetic energy regions
of velocity space. To accomplish this, importance sampling is often
employed \cite{bao2016,aydemir1994}, in which more samples are taken in low probability
regions, and the samples are weighted accordingly. In this case,
expectations on $f$ are computed as weighted averages over the particles.

In a plasma PIC simulation, for example, an accurate representation of the physics
throughout the physical domain may dictate that the marker particles
be evenly distributed throughout physical space, with variable weights
representing the variation of plasma density in space. There may be
similar requirements in velocity space. However, as $f$ evolves in
time, there will be mixing of marker particles that spoils the desired
marker particle distribution, degrading the representation of $f$, and
thus the accuracy of the simulation. Furthermore, in some PIC methods,
to reduce the number of particles and increase the accuracy, a control
variate approach is used \cite{Kleiber2011,hatzky2007,sonnendrucker2015}, 
in which the difference between $f$ and
some reference distribution $f_0$ is represented through marker
particle sampling. In these $\delta f$ methods, the particle weights
evolve, which can lead to sample degeneration in which the weights
become concentrated on just a few particles as the simulation
proceeds; resulting in a poor representation of the distribution. This
phenomena is similar to the sample degeneracy observed in particle
filtering methods \cite{gordon1993, kitagawa1993, kitagawa1996,
douc2005,morelande2011,eyink2006,petetin2013optimal}.

For all of these reasons, it is important that the marker particle
distributions in a PIC simulation be periodically adjusted to maintain
the required importance sampled marker particle distribution. In doing
this resampling, there are conserved physical quantities (e.g. mass,
momentum, energy, angular momentum) that should be preserved throughout
the process. Further the projection of quantities
on to the grid, such as charge and current in a plasma simulation,
should also be preserved through the resampling. In this way, there
will be no change in the field quantities as a result of the resampling.

Many methods for particle resampling that preserve
various moments or distributional features of the solution have been
explored.  Many of these methods rely on splitting and merging the
original particles and are capable of preserving a limited number
of derived features (e.g. moments) to some degree of accuracy.  For
example, Lapenta \cite{Lapenta1994,Lapenta2002,lapenta1995} proposed a scheme in
which the number of particles is increased by splitting particles and
decreased by coalescence of particles close to each other in
phase space. The algorithm is applicable to PIC simulations with
two-dimensional and three-dimensional Cartesian grids and can preserve
overall charge, momentum, and energy.  However, this algorithm cannot
conserve other features of the velocity distribution function, and the
scheme is not directly extendable to 2D or 3D unstructured
grids. Teunissen and Ebert \cite{Teunissen2014} improved Lapenta's
particle merging algorithm using the k--d tree method to search for
the nearest neighbor. Following a similar procedure, Vranic et
al. \cite{Vranic2015} divided the momentum space into smaller
cells for sorting particles that resulted in better local preservation
of the energy, momentum, and charge.

A different two-dimensional method of coalescing particles in PIC
methods that conserves the particle and cell charge and current
densities, as well as the particle energy, is presented by
Assous \cite{Assous2003} . However, the method is limited to
two-dimensional triangular cells and its extension to other cell
geometries and three-dimensions is not straightforward. Moreover, in
this method the number of particles per cell after coalescence is
restricted depending on the integration points employed in the
solution.  Welch et al. \cite{Welch2007} provided an extension of
Assous method \cite{Assous2003} to coalescing particles on 2D and 3D
cells. This method is limited to orthogonal grids and is similar to
the method of Assous, in that coalescence might not be possible in
some cases.

Luu et al. \cite{Luu2016} presented a particle merging algorithm in
which the phase space of a simulation is partitioned into smaller
subsets. The algorithm merges particles that are close to each other
and provides direct control over errors introduced by a merging
event. Examining the performance of this algorithm indicates that
momentum is conserved perfectly while energy conservation is approximate.

Pfeiffer et al. \cite{Pfeiffer2015} proposed two algorithms for
particle splitting and merging that use a 3D unstructured hexahedral
mesh and are expandable to any cell geometry. The first method is
computationally feasible and enables preserving particle charges,
currents, and energies exactly, while the grid projected charge
and current are only preserved approximately. The second
method makes fewer assumptions about the velocity distribution function
resulting in better conservation of the grid quantities, but is
computationally more expensive.


In the work described here, we develop a general algorithm called Moment
Preserving Constrained Resampling (MPCR) for resampling particles in a
PIC simulation, while preserving any desired features of the
distribution. The MPCR algorithm differs from the techniques described
above by using the fact that the marker particles are a sampling
representation of an underlying distribution $f$, and that this
distribution can therefore be resampled when convenient.  To
accomplish this, MPCR uses a binning strategy based on any convenient
discretization of phase space, making it suitable for general
geometries and mesh/grid representations.  It uses constrained
optimization techniques to produce a new set of particle positions,
velocities and weights that preserves any necessary (conditional)
expectations on $f$, including physically conserved quantities and
grid projections, to roundoff error.

The MPCR resampling algorithms described here can be applied to any problem
in which PIC methods are used to represent the evolution of
distributions. But, in this work, the target is a PIC representation
of plasmas, particularly the simulation of edge physics in
tokamak plasmas where particle resampling is most valuable due to the
mixing of marker particles across strong plasma gradients.
More specifically, the target in this case is the
gyrokinetic PIC representation used in the XGC tokamak simulation
codes \cite{KuChang2009_1,sku_2009}. Details of
the MPCR algorithm and its implementation in XGC are described in
section~\ref{sec:method}. A number of example problems demonstrating
the capabilities of MPCR are presented in
section~\ref{sec:results}, and concluding remarks are
provided in section~\ref{sec:conclusions}.

%% file: paper_3_method.tex
In this section, the proposed particle resampling algorithm
is introduced, beginning with the constraints imposed
on the resampled distribution (section~\ref{sec:constraints}), followed
by the resampling process (section~\ref{sec:algorithm}). Finally, as an
example, a description of the implementation of the algorithm for use
with the gyrokinetic code XGC is provided in section~\ref{sec:XGCimp}.

\subsection{Resampling constraints}\label{sec:constraints}
\label{sec:conserved}
\def\bx{\mathbf{x}}
\def\bJ{\mathbf{J}}
\def\bv{\mathbf{v}}
\def\bm{\mathbf{m}}
\def\bzeta{{\boldsymbol{\zeta}}}
\def\avg{\mathrm{avg}}

In the resampling process, we start with a representation of $f$ by a
set of $N^p$ samples (particles) \blue{with specific positions, velocities and weights,} and produce a new representation of
$f$ using a new set of $M^p$ samples. In generating the new samples,
it is important that physically relevant characteristics of the
distribution function be preserved. These include physically important
moments of $f$, as well as the coupling quantities that are projected
from the particle representation to the grid. The general resampling
strategy proposed here allows these quantities to be preserved to high
accuracy. As an example, we consider here a set of constraints that
are relevant to a PIC plasma simulation in which $f(\bx,\bv,t)$
represents the atomic particle distribution in physical space ($\bx$) and
velocity space ($\bv$). In other PIC applications, different constraints would
be appropriate.


First, there are several velocity space moments that are physically
relevant as functions of physical space. Among these are the atomic particle
number density $n$, the momentum density $\bm$, the kinetic energy
density $K$ and the momentum flux tensor (stress tensor) $T$.
\begin{align}\label{eq:f_mom}
n(\mathbf{x},t) & = \int f(\mathbf{x}, \mathbf{v},t) d\mathbf{v},\\
{\bf m}(\mathbf{x},t) & = m \int \mathbf{v} f(\mathbf{x}, \mathbf{v},t) d\mathbf{v},\label{eq:f_momentum}\\
K(\mathbf{x},t) & = \frac{1}{2}m\int |\mathbf{v}|^2 f(\mathbf{x}, \mathbf{v},t) d\mathbf{v},\\
T_{kl}(\mathbf{x},t) & = m\int v_kv_l f(\mathbf{x}, \mathbf{v},t)
d\mathbf{v},
\label{eq:f_stress}
\end{align}
where $m$ is the mass of the atomic particles represented by $f$, and
in (\ref{eq:f_stress}), Cartesian tensor notation is used.  Preserving
some or all of these quantities during resampling avoids introducing
non-physical disruptions into the simulation caused by changes in
these quantities. There are other velocity space moments, which,
depending on the specific application, might also be important.  For
example, in gyrokinetic tokamak plasma simulations, it is often the
canonical angular momentum that is relevant, rather than
$\mathbf{m}$. All of these velocity space moments are functions of
$\bx$, and preserving them as continuous functions of $\bx$ will not
be possible. Instead, they will be preserved in a discrete sense (see
section~\ref{sec:algorithm}), and in a global sense; that is, the integral over the $\bx$
domain will be preserved. These correspond to global conservation of
mass, momentum and energy. In addition, in some applications, it might
be beneficial to preserve configuration space moments (i.e. moments
taken with respect to $\bx$).



For a PIC plasma simulation, the charge density $\rho$ and current
density $\bJ$ are needed as a function of $\bx$ to determine the
electromagnetic fields. These are given trivially by $\rho=qn$ and
$\bJ=q\bm/m$, where $q$ is the charge of the atomic particles
represented by $f$. However, what is important to the calculation is
the representation of $\rho$ and $\bJ$ on the physical space
grid. Consider the projection of a function of $g(\bx)$ on to the discrete
degrees of freedom (e.g. grid point values) defined by
\begin{equation}
g_j=\int\mathcal{P}_j(\bx)g(\bx)\,d\bx
\end{equation}
where $g_j$ is the discrete value for the $j$th degree of freedom, and
$\mathcal{P}_j$ is the projection kernel or shape function, which
depends on the numerical methods employed on the Eulerian grid. In this
case, the discrete values of the charge and current densities $\rho_j$
and $\bJ_j$ are given by:
\begin{align}
\rho_j(t) &= \int\int q \mathcal{P}_j(\bx)f(\mathbf{x}, \mathbf{v},t)\,d\mathbf{v}\,d\bx,\label{eq:charge}\\
\bJ_j(t) &= \int\int q \mathcal{P}_j(\bx)\mathbf{v} f(\mathbf{x}, \mathbf{v},t)\,d\mathbf{v}\,d\bx.\label{eq:current}
\end{align}
To ensure consistency of the electromagnetic fields computed from
$\rho$ and $\bJ$ after the resampling, these grid projections need to
be preserved.

\subsection{The resampling algorithm}\label{sec:algorithm}

To generate a new sample of the distributions function $f$ from an
existing function, the MPCR algorithm begins by discretizing the
sample phase space into bins. The resampling is then done
independently in each bin, and the moments that are to be preserved
are imposed as a constraint on the resampling in each bin.  This
approach has several benefits: 1) any non-ideal or local features of
$f$ (e.g. multi-modality or non-equilibrium peaks) will be preserved
as long as they are resolved by the binning; 2) spatially dependent
velocity moment constraints such as (\ref{eq:f_mom}-\ref{eq:f_stress}) and velocity
dependent configuration moment constraints will be satisfied at the
level of the binning discretization; 3) because the global
characteristics of the distribution are resolved by the binning, the
sampling performed on each bin can be simple, such as sampling from a
uniform distribution on the bin, with weight adjustments to impose
constraints; and finally 4) imposing the moment constraints will be
performed as a constrained optimization, and by doing this on a
bin-by-bin basis one reduces the dimension of each optimization
problem. The details of the algorithm are described below, again using
the case of a kinetic plasma PIC simulation as an example.

\paragraph{Discretizing the phase-space.}
The distribution function $f$ is a function in the six dimensional
physical and velocity space $\bzeta=(\bx,\bv)$. The algorithm begins
by discretizing this six-dimensional space into bins, with each bin
defining a volume in the $\bzeta$ domain. Each of the $N^p$ particles
is a sample of the distribution $f$, and so defines a point in the
six-dimensional $\bzeta$ space, with a given weight. The particles are
sorted into the bins according to their $\bzeta$ coordinate. The
result is that the $N^p$ particles are distributed among $N^b$ bins,
with bin $i$ containing $N^p_i$ particles.  Depending on the
application, it may be convenient to use the spatial discretization of
the grid solver to define the bins.  However, the MPCR can accommodate any
bin geometry.

\paragraph{Defining the number of new particles in each bin.}
Recall, that the purpose of the resampling is to redistribute the
marker particles in $\bzeta$ space to preserve the desired importance
sampling, and if necessary increase or decrease the total number of
particles. The desired importance sampled marker particle distribution
can be defined as $g(\bzeta)$ where $\int_\bzeta g\,d\bzeta=M^p$ is
the target number of particles after resampling. Thus, the target
number of new particles $M^p_i$ in bin $i$ is given by:
\begin{equation}
  M^p_i=\int_{\boldsymbol{\zeta}^i} g\,d\boldsymbol{\zeta},
\label{eq:targetM}
\end{equation}
where $\boldsymbol{\zeta}^i$ is the phase space domain of bin
$i$, i.e. the range of position and velocity coordinates in the bin.

\paragraph{Generating the new samples}
In the $i$th bin, the target number of new samples $M_i^p$ may be
larger or smaller than the old number of samples $N^p_i$. If
$M^p_i\ll N^p_i$ and the variation of the weights among the old
particles is not too great, it is advantageous to draw the $M^p_i$ new
particles as samples from the old particles.  In this case, we use
weighted sampling without replacement \cite{wong1980} 
\blue{
in which a sample in a discrete population has a relative probability to be 
selected according to its weight and, once selected, is removed from the population. 
This method enables drawing
}
$M^p_i$ new samples while precluding the possibility of an old
particle being sampled multiple times. This is important when the
subsequent particle evolution is deterministic. Provided that
$M^p_i/N^p_i$ is sufficiently small compared to
$\avg_j(w^i_j)/\max_j(w^i_j)$, where $w^i_j$ is the weight of the
$j$th particle in bin $i$, the probability of $M^p_i$ random draws from
the old weighted particle distribution including any particle more
than once is small. When this is so, uniformly weighted samples
obtained using the sampling without replacement algorithm approach an
unbiased sample from $f$ on the bin, which is what makes this approach
advantageous.

When this condition is not satisfied, a completely new set of $M^p_i$
particles is sampled from a convenient distribution on the bin domain
$\bzeta^i$. A uniform distribution on $\bzeta^i$ is used here because
it is \blue{the maximum entropy distribution supported in an interval (i.e., bin)}, and has the
advantage of being simple. A maximum entropy distribution constrained
by some or all of the moment constraints is another possibility, but
with sufficiently small bins this is of little value while being
significantly more complicated. 

\paragraph{Imposing constraints}
Within the bin, the weights of the new particles are intended to be
uniform. There are two reasons for this. First, when the new particles
are sampled from the old particles, the uniformly weighted particles
approach a sampling of $f$. Second, for a given number of samples,
uniformly weighted samples provide the most information about the
distribution. However, in general, regardless of how the new samples
are drawn, the samples will not satisfy the constraints as described
in section~\ref{sec:constraints}. To impose these constraints, the weights are
adjusted, while keeping them as close to uniform as possible through a
constrained optimization.

To impose the constraints, the moment integrals are first restricted
to the bin domain and then evaluated as a sum over the samples. For
example, the bin restricted momentum from (\ref{eq:f_momentum}) is given by
\begin{equation}
\blue{
\bm^i=m\int_{\boldsymbol{\zeta}^i} \bv f(\bzeta)\,d\bzeta\approx
m\sum_{j=1}^{N^p_i}\bv^i_j w^i_j
}
\end{equation}
where $\bm^i$ is the average momentum in bin $i$ and $\bv^i_j$ is the
velocity of particle $j$ in bin $i$. Similar approximations of
bin-wise atomic particle count $n^i$, kinetic energy $K^i$ and
momentum flux tensor $T_{kl}^i$ can easily be written from equations
(\ref{eq:f_mom}--\ref{eq:f_stress}). The objective is to ensure that these bin-wise
quantities are the same when computed with the new samples as with the
old samples. In addition, we impose similar constraints on the first
and second configuration moments, primarily to improve the accuracy of
the representation of $f$ on the bin. The constraints to be imposed
through adjustment of the weights $\tilde w_j^i$ on the new particles
are thus:
\begin{align}
  \sum_{j=1}^{M^p_i}\tilde w^i_j &=\sum_{j=1}^{N^p_i}w^i_j,\label{eq:0const}\\
  \sum_{j=1}^{M^p_i}\tilde{\boldsymbol{\zeta}}^i_j\tilde w^i_j &=\sum_{j=1}^{N^p_i}\boldsymbol{\zeta}^i_j w^i_j,\label{eq:1const}\\
  \sum_{j=1}^{M^p_i} (\tilde{x}_k \tilde{x}_l)^i_j\tilde w^i_j &=\sum_{j=1}^{N^p_i}(x_k x_l)^i_jw^i_j,\label{eq:2xconst}\\
  \sum_{j=1}^{M^p_i} (\tilde{v}_k \tilde{v}_l)^i_j\tilde w^i_j &=\sum_{j=1}^{N^p_i}(v_k v_l)^i_jw^i_j,\label{eq:2vconst}
\end{align}
where $\tilde\bzeta^i_j$, $\tilde\bx^i_j$ and $\tilde\bv^i_j$ are the
values of $\bzeta$, $\bx$ and $\bv$ for $j$th new particle 
in bin $i$. 
\blue{
Thus, in addition to preserving the physical quantities $n^i$, $\bm^i$, $K^i$
and $T_{kl}^i$ as a consequence of (\ref{eq:0const}) - (\ref{eq:2vconst}), the 
conservation of moments in configuration space,
$
\int \mathbf{x} f(\mathbf{x}, \mathbf{v},t) d\mathbf{x}
$
and
$
\int \mathbf{x}^2 f(\mathbf{x}, \mathbf{v},t) d\mathbf{x}
$,
} lead to a better bin-wise representation of $f$

Similarly, the grid projection quantities (charge and current
densities in the plasma example) can be defined bin-wise, and the
projection integrals as approximated by the sampling of $f$ preserved
through the resampling. For example, the current density projection
(\ref{eq:current}) can be defined bin-wise as:
\begin{equation}
\bJ^i_m=q\int_{\bzeta^i}\mathcal{P}_m(\bx)\bv
f(\bzeta)\,d\bzeta\approx q\sum_{j=1}^{N^p_i}\mathcal{P}_m(\bx^i_j)\bv^i_j
w^i_j,
\end{equation}
for each $m$ for which bin $i$ intersects the support of $\mathcal{P}_m$.
Here $\bJ^i_m$ is the contribution of bin $i$ to degree of freedom
$m$ of the grid representation of the current.
Therefore to preserve the grid projection of charge and current density,
the following constraints are imposed:
\def\bin{\mathrm{bin}}
\newcommand{\supp}{\mathop{\mathrm{supp}}}
\begin{align}
\sum_{j=1}^{M^p_i}\mathcal{P}_m(\tilde\bx^i_j)
\tilde w^i_j &= \sum_{j=1}^{N^p_i}\mathcal{P}_m(\bx^i_j)w^i_j
\qquad\forall\, m \ni \bin_i\cap \supp(\mathcal{P}_m)\neq\emptyset
\label{eq:chargeconst}
\\
\sum_{j=1}^{M^p_i}\mathcal{P}_m(\tilde\bx^i_j)\tilde\bv^i_j
\tilde w^i_j &= \sum_{j=1}^{N^p_i}\mathcal{P}_m(\bx^i_j)\bv^i_jw^i_j
\qquad\forall\, m \ni \bin_i\cap \supp(\mathcal{P}_m)\neq\emptyset
\label{eq:currentconst}
\end{align}
where $\bin_i$ is the configuration space domain of bin $i$.

The weights $\tilde w^i_j$ in the new sample can now be set to ensure that
the constraints (\ref{eq:0const}-\ref{eq:2vconst}) and
(\ref{eq:chargeconst}-\ref{eq:currentconst}) are satisfied,
while minimizing the variance in the weights by solving the
following optimization problem:
\begin{equation}\label{eq:quadprog}
  \{ \tilde{w}^i_j \}_{(j\in 1,2,\ldots M^p_i)}=\argmin_{\tilde{w}^i_j\in \mathcal{C}^i , \; \tilde{w}^i_j>0} \sum_{j=1}^{M^p_i}(\overline{\tilde
    w}^i-\tilde w^i_j)^2,
\end{equation}
where
\begin{equation}
\blue{
  \overline{\tilde w}^i=\frac{1}{M_i^p}\sum_{j=1}^{N^p_i} {\tilde w}^i_j,
}
\end{equation}
and $\mathcal{C}^i$ is the set of $\tilde w^i$ satisfying the
constraints (\ref{eq:0const}-\ref{eq:currentconst}). Notice that with
a quadratic objective function (\ref{eq:quadprog}) and linear equality
constraints in $\tilde w^i_j$ (\ref{eq:0const}-\ref{eq:currentconst}),
this is a straight-forward quadratic programming problem, for which
efficient and well established algorithms are available \cite{trove.nla.gov.au/work/21391104}.

\paragraph{Comments, Remarks, and Discussion } 

The algorithm described here yields a resampling of the
distribution $f$ by discretizing the phase space into bins, with the
assumption that $f$ will then vary little over each bin. That is, that
the binning provides an accurate discrete representation of
$f$. 
\blue{Clearly, the smaller the bin size, the more accurately the
binning can represent $f$, but also the fewer particles will be in
each bin. However, the number of particles in each bin must be
greater than the number of constraints to be imposed during the
resampling, preferably significantly greater. There is therefore a potential tradeoff in accuracy
between decreasing bin size and increasing the number of constraints
(e.g. imposing constraints on higher order moments), which is similar
to the tradeoff between decreasing element size and increasing element
order in finite element methods. In practice, as discussed
in \S\ref{sec:XGCimp}, the configuration space is binned according to the
underlying spatial grid cells, and velocity space has been binned
on a (typically) 32$\times$32 rectangular array.  
An adaptive approach to determining
adequate bin sizes based on estimates of the resampling error (e.g. in
unconstrained moments) may be useful.
%
}

\blue{
The constrained optimization to minimize weight variances within a
bin, as expressed in (\ref{eq:quadprog}), is similar in formulation to
the use of the \textit{principle of maximum uniformity} proposed by Lapenta \&
Brackbill \cite{lapenta1995}. However, because here we are concerned
with a sample-based representation of a distribution function ($f$),
the motivation is quite different. By minimizing the variance of the
particle weights, we ensure that the particles carry near-maximum
information about the distribution; whereas, in \cite{lapenta1995},
the objective is to ensure a smooth representation of a continuous
field. The motivation here is more closely related to the use of uniformly
weighted samples in Monte Carlo methods.
%
}

In some particle evolution models, such as $\delta f$ formulations as
described in section~\ref{sec:intro}, the particle weights must
evolve. This can lead to sample degeneration in which weights become
concentrated on fewer and fewer particles as the simulation proceeds,
resulting in a poor distributional representation. Minimizing particle
weight variance (\ref{eq:quadprog}) in the optimization step of the
resampling algorithm alleviates this degeneracy. As in particle
filtering methods, maintaining relatively uniform particle weights
improves the representation of the distribution, see
e.g. \cite{beevers2007}.

MPCR has three notable features compared to recently developed
particle splitting/merging algorithms in the PIC simulation
context:

\begin {enumerate}

\item The sampling strategies used to define new particle
distributions is easily applied to bins with irregular geometries,
including, for example, unstructured meshes. This enables straight
forward integration of the algorithm in any available PIC code.
Moreover, since the optimization problems solved on each bin are
independent, the computations at the bin level can be easily
parallelized or built into the parallel solver of the PIC code.

\item As will be shown in the numerical results section,
the constrained optimization technique is able to conserve any number
of particle and grid quantities to \textit{near machine precision}
regardless of the method employed in the PIC code for particle
evolution and deposition. Some of the previously developed algorithms
in the literature \cite{Lapenta2002, timokhin2010, nerush2011} are
successful in preserving some quantities but fail to preserve
grid projections or other characteristics of the distribution
function.  For example, the particle
merging algorithm developed by Luu et al. \cite{Luu2016}
preserves  total energy to
only $10^{-3}$.
In addition, the
statistical particle split and merge methods recently proposed by
Pfeiffer et al. \cite{Pfeiffer2015}, preserve grid projections to
machine precision (or as reported to $10^{-19}$) for the low-order
cell mean projection, but is much less accurate (only $10^{-2}$) for
higher order grid projections.

\item The bin-wise constrained optimization used in the resampling
algorithm can introduce a subtle parallel inefficiency. The problem
arises when there are too few particles in a bin to satisfy all the
constraints. This is easily addressed by merging neighboring bins
until there are enough particles in the merged bin. This merging
process disrupts the parallel efficiency of the optimization
calculations. This is not a serious concern because resampling is done
infrequently, so that the cost of resampling including the extra cost
of bin merging, is small compared to time evolving the PIC simulation.
Marker particle weight distributions can be used to adaptively
determine when resampling is necessary. 

\end{enumerate}

\subsection{Implementation in the XGC gyrokinetic PIC code}
\label{sec:XGCimp}

MPCR can be applied to resample any distribution function, but as an
example application we apply it here to a gyrokinetic plasma PIC
code.  In this section, we describe the XGC gyrokinetic PIC
codes for tokamak fusion reactor
simulations \cite{KuEtal_2018,KuChang2009_2} along with  specific
implementation issues for integrating MPCR.

The time evolution of plasma systems is described by the six
dimensional Maxwell-Boltzmann system in phase
space \cite{wesson2011tokamaks}.  In strongly magnetized plasmas, such
as tokamak fusion plasmas, these equations can be averaged
analytically over the gyrophase
(viz. the cyclotron frequency) leading to a gyrocenter tracking
equation for a charged ring that evolves with the relatively slow motion of
particle gyrocenters. This treats rapid particle orbits about
magnetic field lines asymptotically.  Transforming to the gyrocenter
coordinate then results in a phase space reduction from six to five
dimensions, and the resulting system of equations is referred to as
the gyrokinetic equations. While a significant simplification is
achieved through the reduction of the full six-dimensional equations
to five-dimensional gyrokinetic equations \cite{hahm1988}, simulating
this five-dimensional system in a full-scale tokamak fusion reactor
is still a formidable task requiring a carefully formulated
numerical scheme and large-scale high-performance computers.


The Lagrangian evolution equations for the marker particle positions,
velocities and weights depend on exactly how $f$ is sampled. In a
straight-forward approach (called full-$f$), the distribution is
simply sampled. However, this requires the largest number of particles
to accurately represent $f$, and is therefore computationally most
expensive. Other algorithms use a control-variate ($\delta f$)
approach in which the difference between $f$ and an ideal or simple
distribution $f_0$ is sampled. There are a number of such $\delta f$
approaches that differ in the details; see for
example \cite{Ku2016467}. In the full-$f$ representation, the particle
weights do not evolve in time, but in $\delta f$ approaches, they
do. The details of a resampling algorithm for a full-$f$ and $\delta
f$ sampling scheme may differ, because of the difference in
representation. In the algorithm described here, we consider a
full-$f$ sampling representation of $f$ for simplicity.

\begin{figure}[htp]
\begin{center}
  \includegraphics[width=0.5\textwidth,clip]{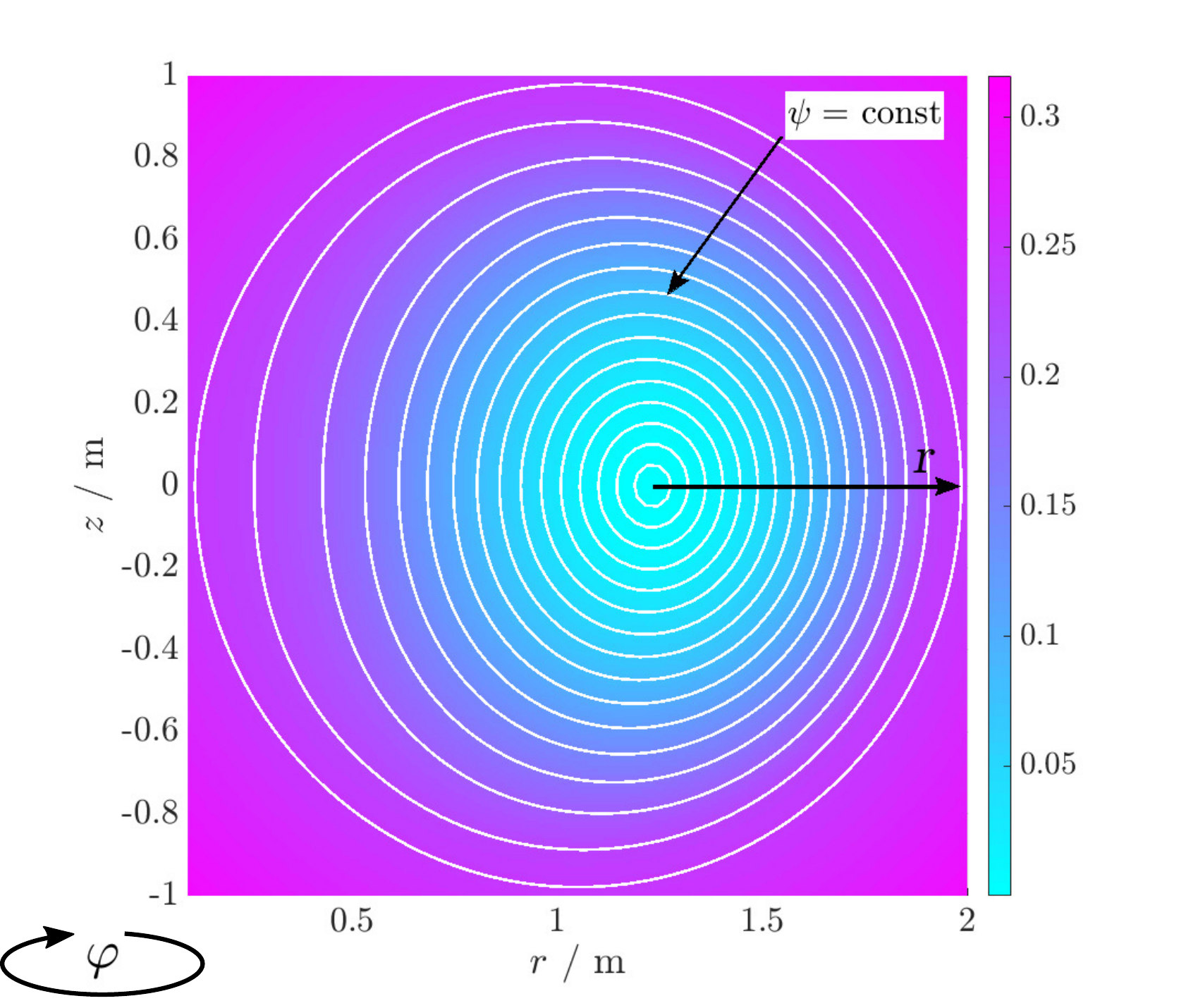}
\end{center}
\caption{\blue{Equipotentials of the flux function $\psi$, along with the cylindrical coordinate axes used in the XGC code.}} \label{fig:coordax}

\end{figure}

The XGC code-base supports a large collection of PIC solution
strategies, allowing for full-$f$, delta-$f$, and total-$f$ (or
equivalently, hybrid-Lagrangian \cite{Ku2016467}) simulations, where
in each case the XGC PIC algorithms support turbulence over a plasma
volume parameterized in a toroidal reactor geometry, following the
magnetic axis across the magnetic separatrix and scrape-off layer
(SOL), to just outside the sheath interfacing material boundary.  The
XGC PIC model uses a cylindrical coordinate system, in which the
components of the particle position vector are $\mathbf{x} =
(r,z,\phi)$, and the velocity vector $\mathbf{v}$ decomposes into parallel and
perpendicular components along the magnetic field $\mathbf{B}(r,z,\phi)$
via projections $v_\parallel=\mathbf{v}\cdot\mathbf{B}/|\mathbf{B}|$ and $v_\perp=|\mathbf{v}-v_\parallel\mathbf{B}/|\mathbf{B}||$.
The code propagates marker particles using
Lagrangian motion and the corresponding fields are solved on a finite
element mesh. Due to the need to represent irregular material
boundaries and the magnetic X-point, XGC uses a 2D unstructured
triangular mesh \cite{adams2009scaling}. A constrain imposed on the mesh is
that nodes must lie on magnetic flux surfaces of constant $\psi$ (see Fig.~\ref{fig:coordax}), which also acts as an effective radial coordinate for 1D
diagnostics.

\begin{figure}[tp]
\centering
	\includegraphics[trim = 100mm 20mm 100mm 20mm, clip, width=.4\textwidth]{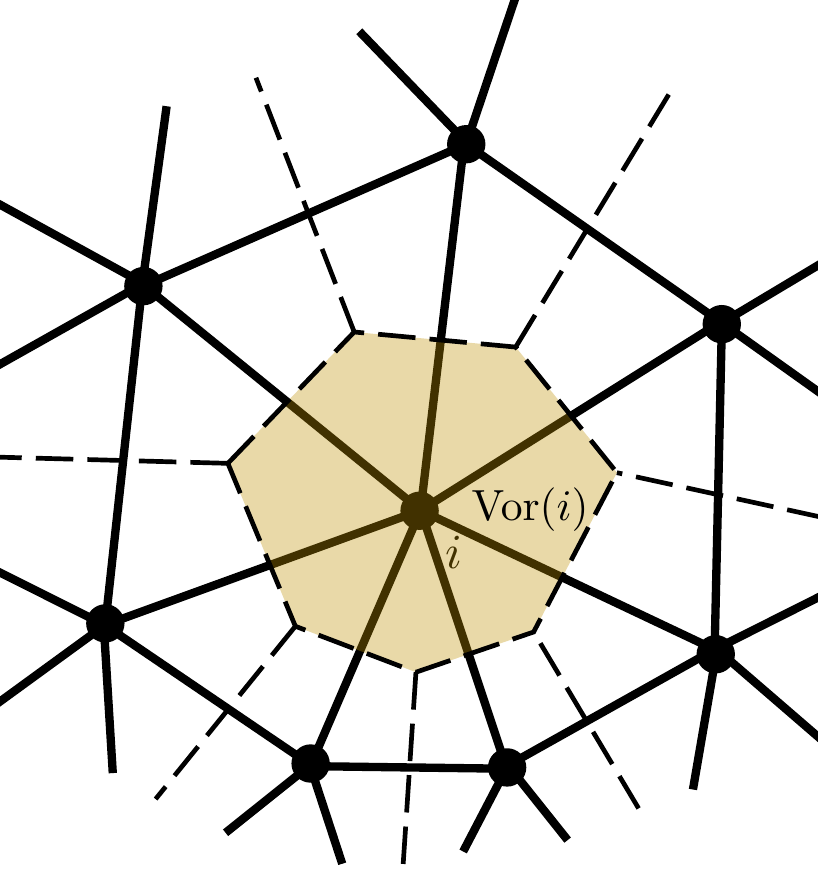}
	\\ \vspace{-0.1in}
 	\caption{Illustration of a 2D unstructured mesh and Voronoi cell associated with the vertex (node) $i$.
	\blue{The finite element nodes and elements are represented by points and solid lines.
	The dashed lines indicate the Voronoi cells.}
	}	
	\label{fig:voronoi}
\end{figure}

Integrating MPCR into the XGC PIC algorithm begins with
the Eulerian spatial discretization, which is based on unstructured
triangular meshes in planes of constant $\phi$.
In XGC, for the purpose of parallel computation, the particles are
sorted into specific sub-volumes of the domain defined through the
mesh, and we use these volumes as the configuration-space MPCR bins to
avoid additional parallel communication. These volumes are defined by
considering the two-dimensional $r$-$z$ mesh at the points in $\phi$
that are half-way between the constant-$\phi$ planes of the Eulerian
grid. The plane is tiled into \blue{\textit{Voronoi cells}} around each vertex
of the two-dimensional mesh, as shown in figure~\ref{fig:voronoi}. The binning
volumes are then defined as the extrusion of the Voronoi cells along
the magnetic field lines to the closest constant-$\phi$ planes of the
grid. Sorting into these Voronoi bins effectively groups together the
particles closest to the same constant-$\phi$ mid-plane whose
projection along the field lines onto that mid-plane is closest to the
same mesh vertex \cite{Dwyer1991}. In the MPCR algorithm, the
particles in each Voronoi bin are further sorted into uniformly spaced
Cartesian bins in the two-dimensional gyrokinetic velocity space
$\mathbf{v} = (v_\parallel ,v_\perp)$, and moment-constrained
resampling is performed on each Voronoi/Cartesian bin.

\def\Vertex{\mathrm{Vert}}
The resampling constraints include both simple moment constraints such
as (\ref{eq:0const})-(\ref{eq:2vconst}) and the grid-projection
constraints (\ref{eq:chargeconst})-(\ref{eq:currentconst}). The
definition of the moment constraints is straight-forward, and the
operators $\mathcal{P}_m$ in the grid-projection constraints are
defined based on the numerical operators used in XGC to project the
particle information onto the Eulerian grid. Specifically, the
Eulerian degrees of freedom $m$ are defined on the vertices of the
three-dimensional grid, which is more naturally enumerated by the
index $k$ of the vertex on the two-dimensional triangular mesh, and
the index $l$ of the constant-$\phi$ grid plane. For any point in
configuration space $\bx$, let $\triangle^+(\bx)$ be the mesh triangle
in which the magnetic field line through $\bx$ intersects the closest
grid plane in the positive $\phi$ direction, enumerate this grid plane
as $l^+(\bx)$, and let $\Vertex(\triangle^+(\bx))$ be the set of three
vertices of $\triangle^+(\bx)$. Further, let $\lambda_k^+(\bx)$ for
$k\in \Vertex(\triangle^+(\bx))$ be the barycentric coordinate in
$\triangle^+(\bx)$ associate with vertex $k$ of the magnetic field
line projection of $\bx$. Finally let $\beta^+(\bx)$ be the distance
along the field line from $\bx$ to plane $l^+(\bx)$ normalized by
distance along the field line from plane $l^+(\bx)$ to plane
$l^-(\bx)$, which is the closest constant-$\phi$ plane to $\bx$ in the
negative $\phi$ direction. Similarly $\triangle^-$, $\lambda_k^-$ and
$\beta^-$ are the same for the grid plane $l^-(\bx)$. Note that
$\beta^+(\bx)+\beta^-(\bx)=1$ and $\sum_k\lambda_k^\pm=1$, where the
sum is over $k\in\Vertex(\triangle^\pm(\bx))$.  Then the grid
projection operator $\mathcal{P}_{k,l}(\bx)$ is given by
\begin{equation}
\mathcal{P}_{k,l}(\bx)=
\begin{cases}
\beta^+(\bx)\lambda^+_k(\bx)&\mbox{if $k\in\Vertex(\triangle^+(\bx))$
and $l=l^+(\bx)$}\\
\beta^-(\bx)\lambda^-_k(\bx)&\mbox{if $k\in\Vertex(\triangle^-(\bx))$
and $l=l^-(\bx)$}\\
0&\mbox{otherwise}
\end{cases}
\end{equation}
This then completes the definition of the charge and current grid-projection
constraints in (\ref{eq:chargeconst})-(\ref{eq:currentconst}).

%% file: paper_4_results.tex

In this section we present several numerical examples motivated by the
use of MPCR in {full-$f$ gyrokinetic simulations}. In these
examples, the constrained optimization problem (\ref{eq:quadprog}) is
solved using the {well-established} algorithm of Goldfarb \& Idnani \cite{goldfarb1983}
as implemented in the \textit{quadprog} library \cite{quadprog}. 
{There are numerous other optimization algorithms that could
be used for this straight-forward quadratic programming problem.}

In the first set of test problems, we demonstrate the ability of MPCR
to preserve features of the distribution function when resampling an
arbitrary distribution and resampling particles from an XGC plasma
turbulence solution. Finally we show the benefits of periodic particle
resampling using MPCR on a neoclassical tokamak plasma simulation.  {In a
neoclassical simulation, transport is represented by a diffusion
process that is formulated to account for the effects of the magnetic
geometry, which may produce complex particle orbits and drifts.}

{In addition to the tests reported here, the resampling algorithm was
applied to Landau damping simulations to ensure that resampling
preserves the phase-space structure of the solution, as expected it
does. However, Landau damping is not a compelling case for the use of
the resampling algorithm {described here} because it does not require
strong particle weight variations. It will therefore not be considered
further here. {This is in contrast to \emph{remapping} algorithms
applicable to a different class of PIC methods as described
by \cite{doi:10.1137/16M105962X}, which have the objective of
preserving the placement of computational particles on a Cartesian
mesh in physical and velocity space, and which \emph{are} of value in
the Landau damping problem.} }

\subsection{Global resampling examples}

%
%

\begin{figure}[tp]
\centering
	\includegraphics[trim = 140mm 0mm 140mm 0mm, clip,width=.48\textwidth]{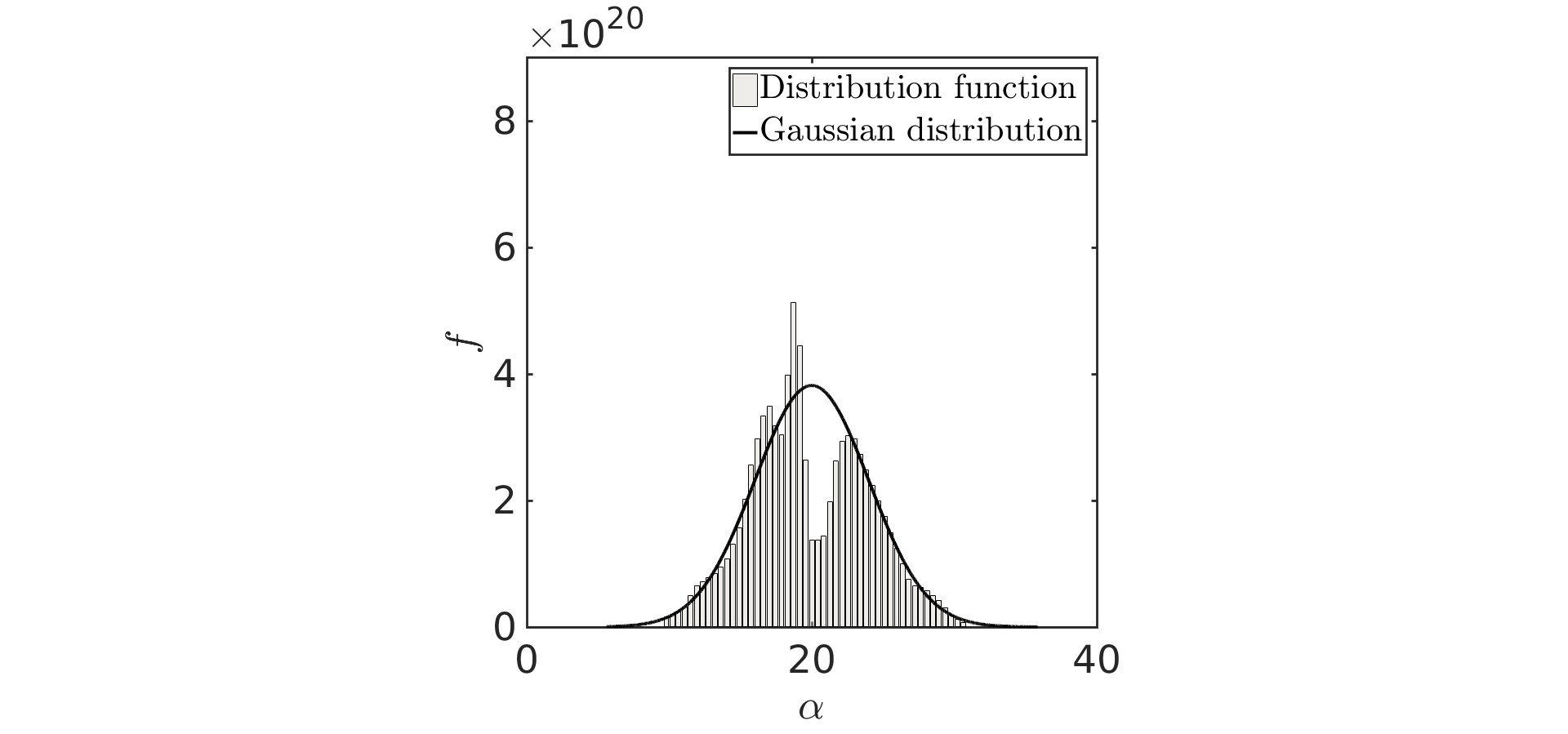}
        ~
        \includegraphics[trim = 140mm 0mm 140mm 0mm, clip,width=.48\textwidth]{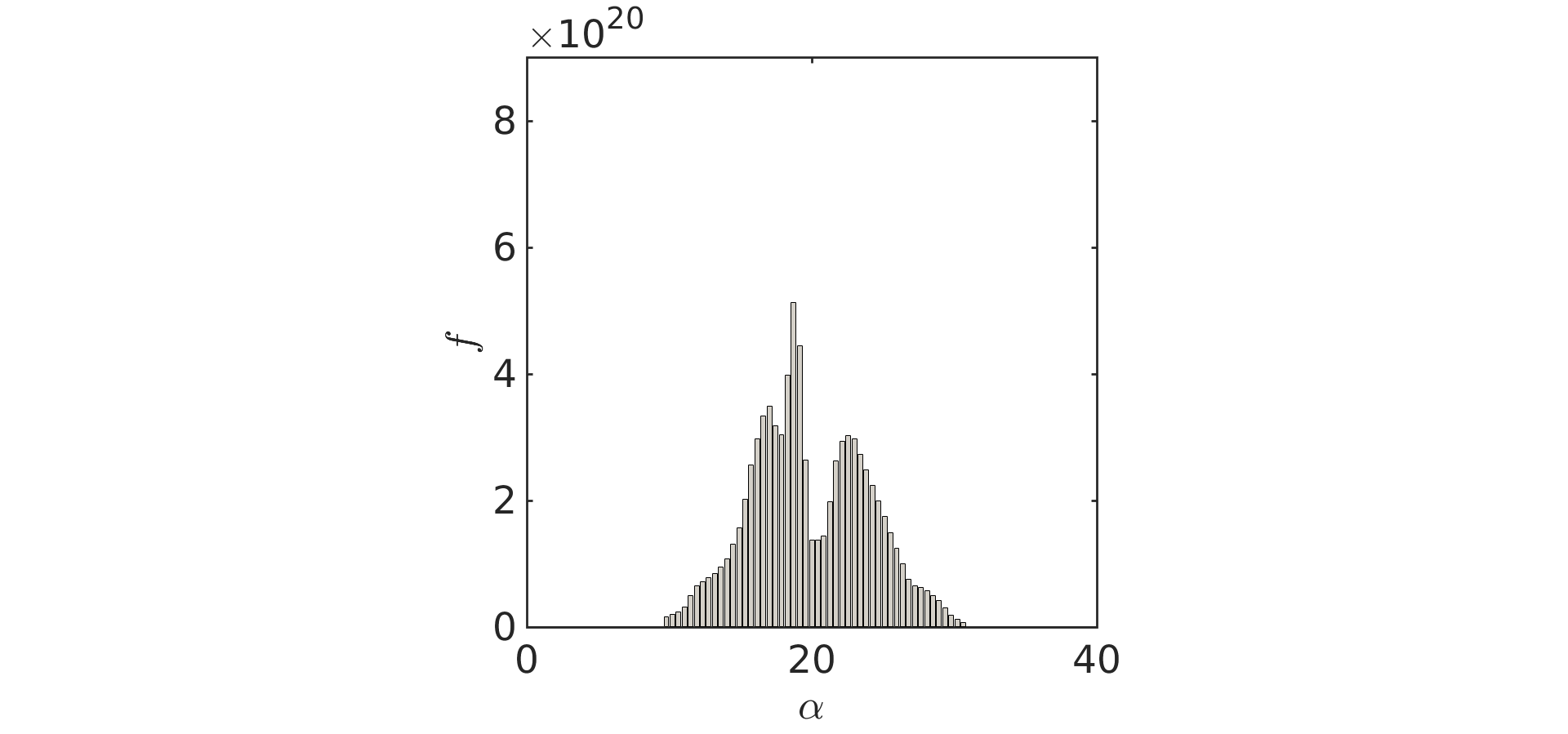}
	\\ \hspace{0.0in} (a) \hspace{2.6in}
	  (b) \\
          \includegraphics[trim = 140mm 0mm 140mm 0mm, clip,width=.48\textwidth]{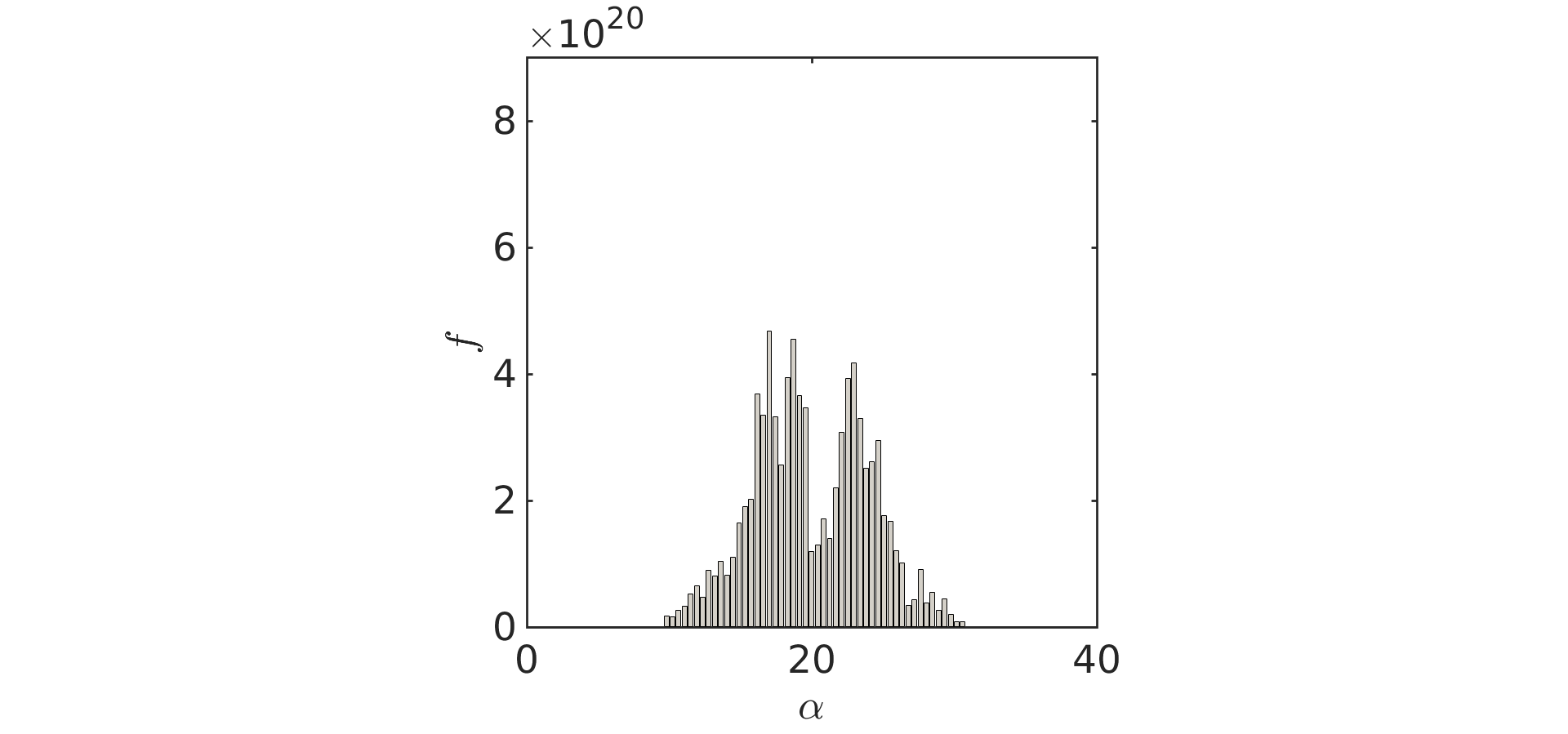}
	  ~
          \includegraphics[trim = 140mm 0mm 140mm 0mm, clip,width=.48 \textwidth]{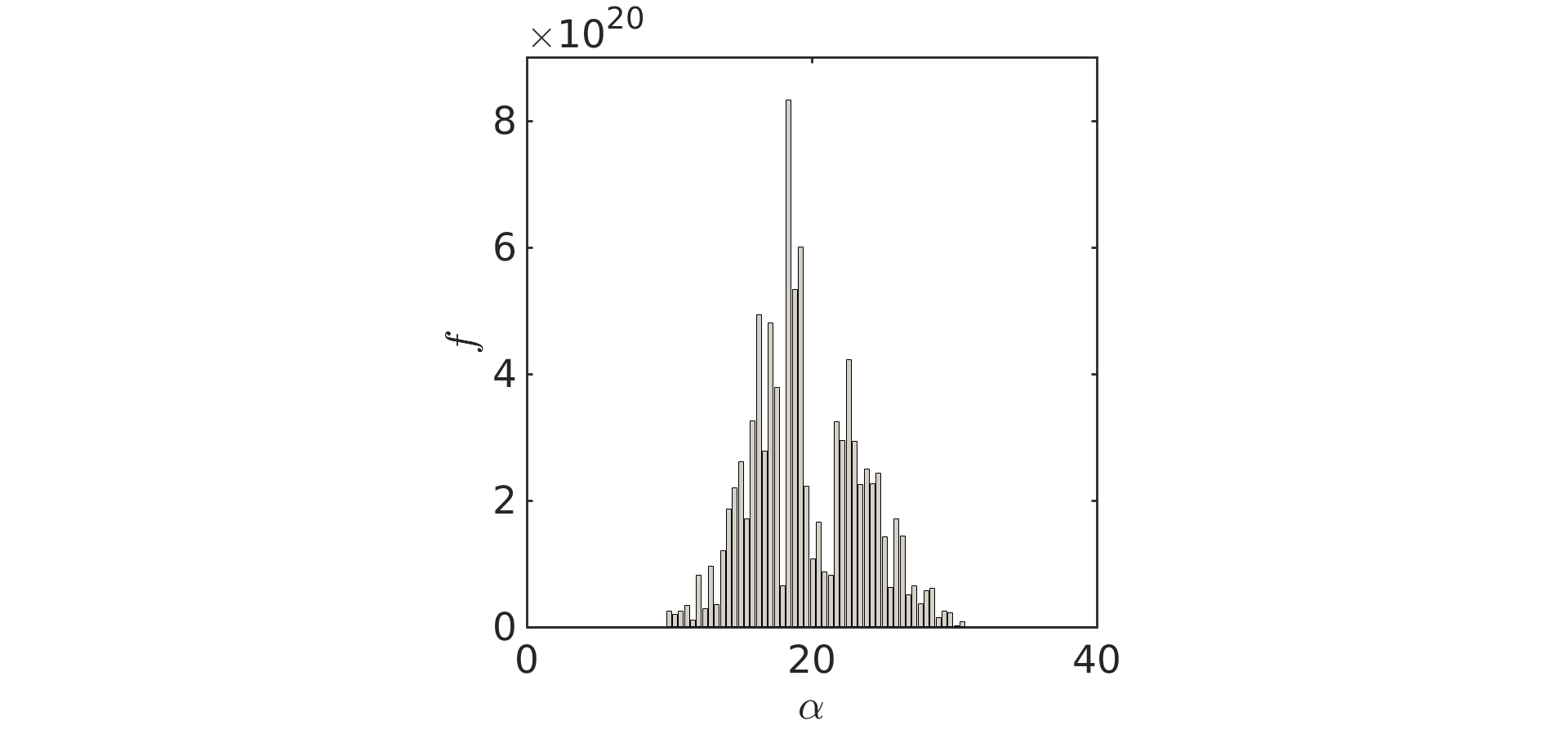} \\ \hspace{0.0in}
	  (c) \hspace{2.6in} (d) \vspace{-0.1in}
          \caption{Synthetic
	  distribution $f$ \blue{as a function of general phase space
        variable $\alpha$}, as represented initially by $N^p=50000$
	  marker particles, and down-sampled by a factor 50 and 200,
	  using MPCR and global random down-sampling. The MPCR algorithm is
	  applied with 50 bins, which are the same as the bins used
	  in the histograms. Shown are:
          (a) initial distribution,
          (b) MPCR with $\mathrm{DSF}=200$ 
          , and global random down-sampling with (c) $\mathrm{DSF}=50$ and
          (d) $\mathrm{DSF}=200$. In (a), the curve is a Gaussian.}	
	\label{fig:1d_localfeature_dsf}
\end{figure}
%

For the first example, consider the one-dimensional probability
distribution represented by the histogram shown in
Figure \ref{fig:1d_localfeature_dsf}(a). It might, for example, be a
distorted marginal distribution of one of the velocity components. In
this case, Maxwell-Boltzmann statistics would yield a Gaussian
distribution, and the given distribution is clearly far from Gaussian.
\blue{
In this figure, the synthetic distribution function $f$ is presented
as a function of a general phase space variable $\alpha$,
which could, for example, be any velocity or position coordinate of the phase space.
}
This distribution is represented with $N^p=50000$ samples \blue{with non-uniform weights generated from a perturbed Gaussian,
and down-sampled to $M^p=1000$ and $M^p=250$: down-sampling factors (DSF)
of 50 and 200, respectively.}

\begin{figure}[tp]
\centering
	\includegraphics[trim = 20mm 0mm 40mm 0mm, clip, width=.3\textwidth]{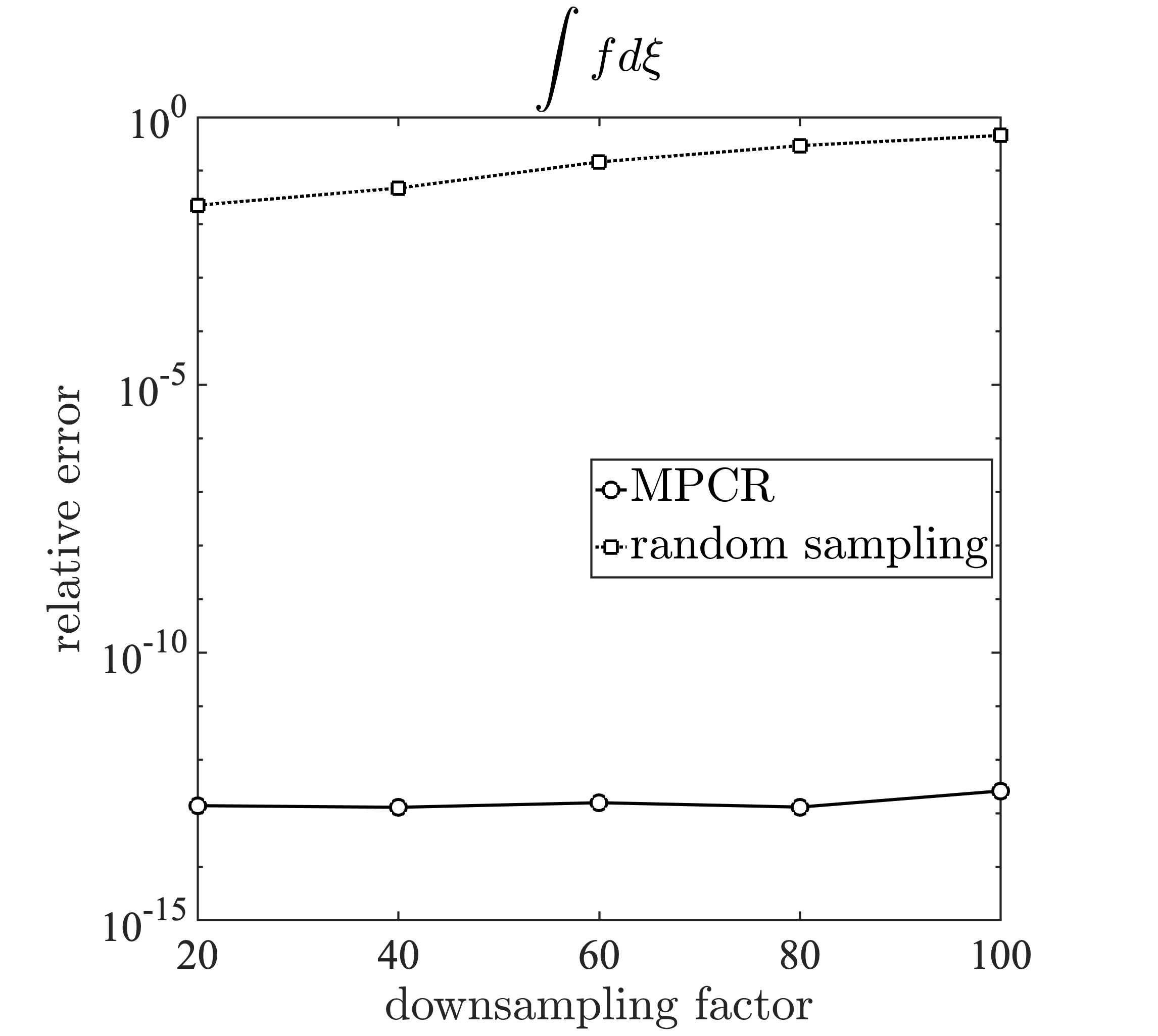}
	~
	\includegraphics[trim = 20mm 0mm 40mm 0mm, clip, width=.3\textwidth]{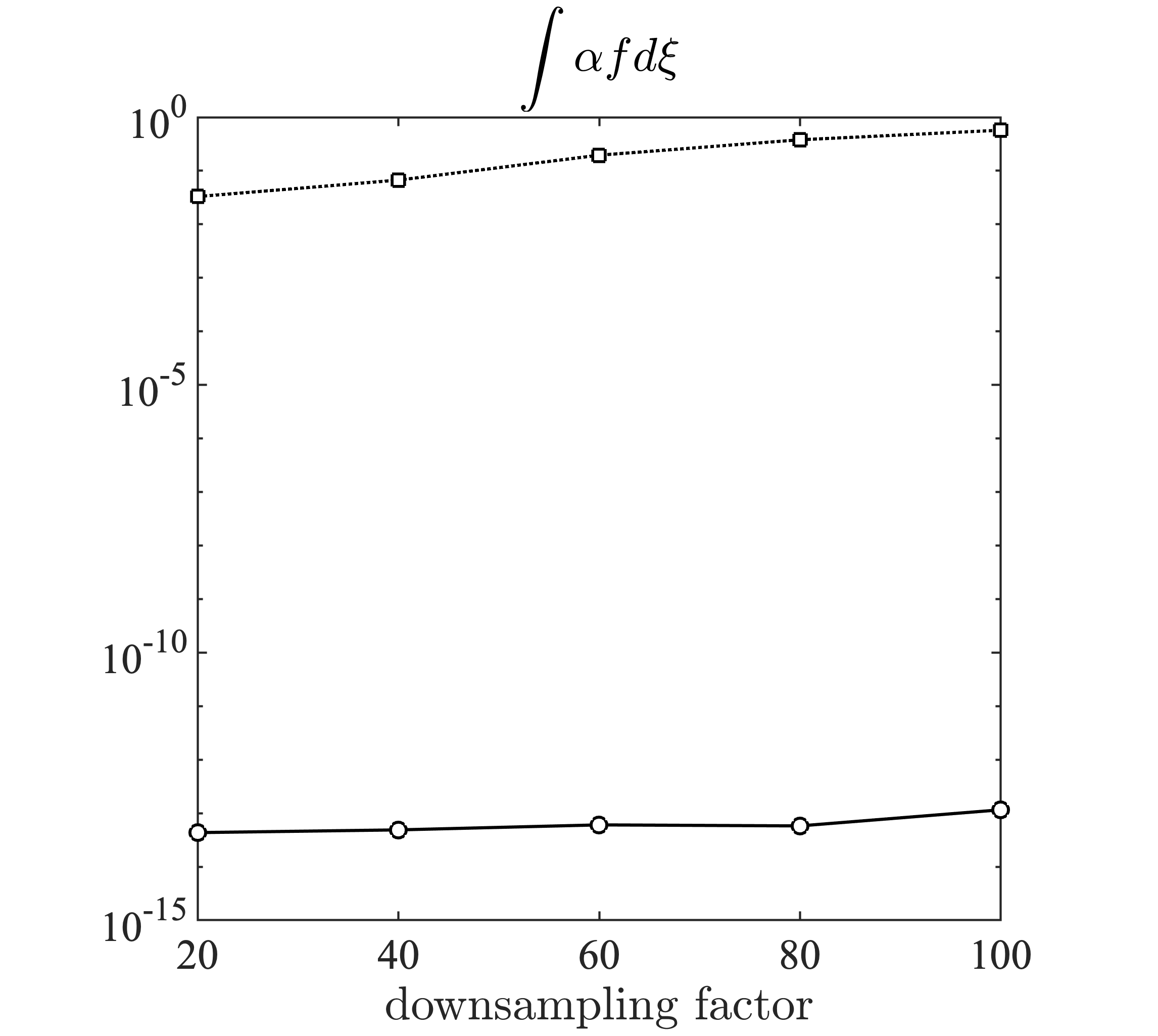}
	~
	\includegraphics[trim = 20mm 0mm 40mm 0mm, clip, width=.3\textwidth]{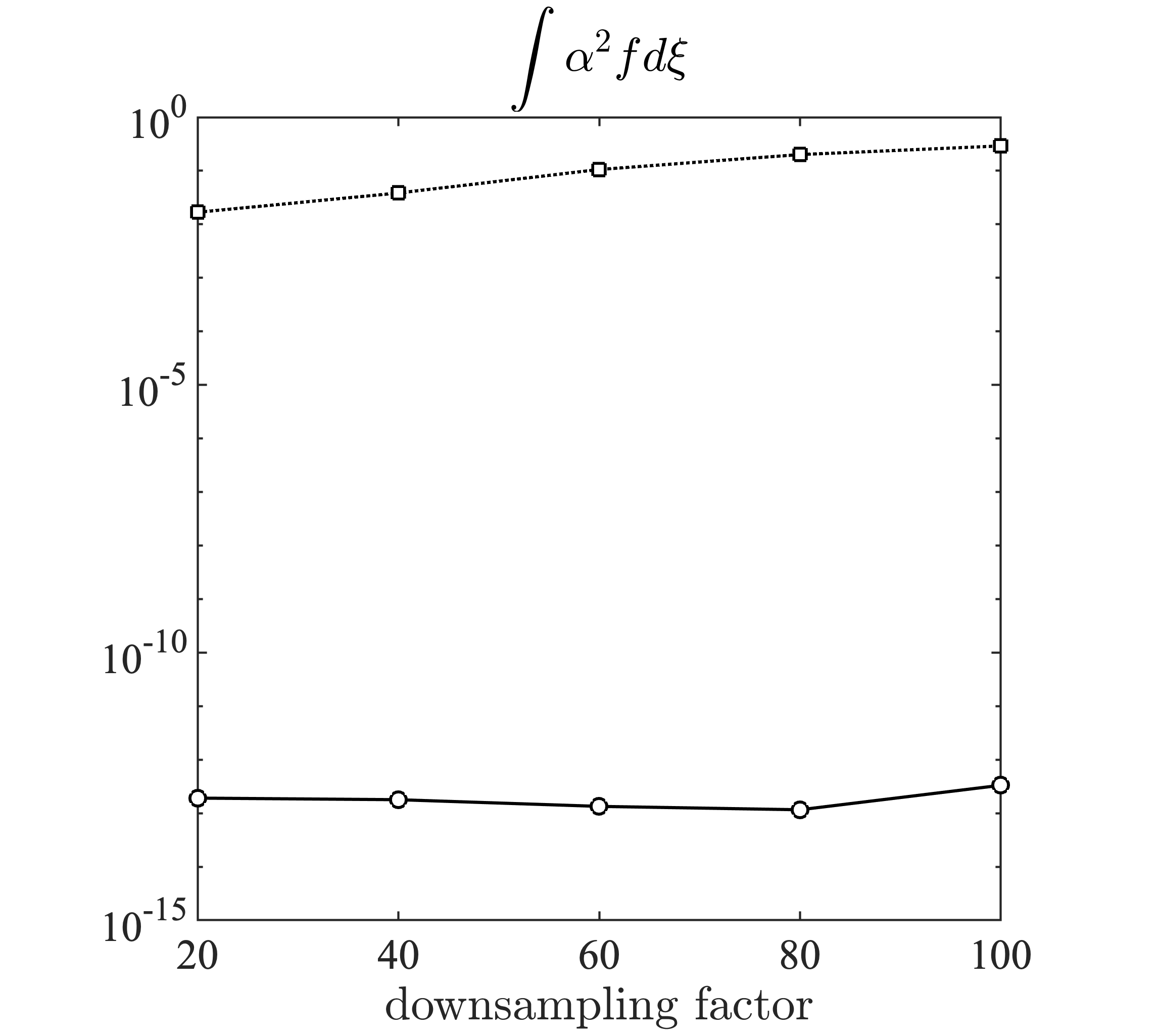}	
	\caption{Relative error in global zeroth, first, and \blue{second} moments due to feature-based/random down-sampling for different down-sampling factors
	(error in random down-sampling is the mean of $10^4$
	realizations).}
	\label{fig:1d_error}
\end{figure}

Down-sampling was performed using MPCR in one-dimension, using
$M^b=50$ bins in which the sample density, and first and second
moments are preserved.  For comparison, global random down-sampling
was also used\blue{, in which $M^p$ new marker particles are selected randomly from the $N^p$ old 
particles. Such global down-sampling has been used in the past in
plasma PIC codes, for example, in the
XGC plasma simulation codes.}
The results are shown in
Figures \ref{fig:1d_localfeature_dsf} and \ref{fig:1d_error}. By
construction, the histogram formed on the 50 down-sampling bins is
identical to that of the original sample, but the randomly down-sampled
histograms show large variations from the original.  Indeed, the
sampling error obscures the structure of the distribution; so much so
that in the DSF=200 case, the underlying structure of the distribution
is not visible at all. By construction, the low-order moments of the
distribution are preserved in MPCR, as shown in
Figure \ref{fig:1d_error}, whereas relative errors in the moments with
random down-sampling range from 2\% to 10\%. 
%
%
%

MPCR up-sampling was also tested on the distribution function from
Figure \ref{fig:1d_localfeature_dsf}(a), with similar results. That is,
constrained moments are preserved to roundoff error, while detailed features
of the distribution are preserved on the scale of the binning (not
shown). A common alternative up-sampling approach is particle
reproduction \cite{douc2005}, in which each particle is duplicated $n_u$
times, where $n_u$ is the up-sampling factor. This will obviously
preserve all moments and all features of the sampled distribution
function. However, if the subsequent evolution of the particles in the
PIC code is deterministic, it will accomplish nothing, since the
particles will remain identical for all time and not provide an
improved sample of $f$. To avoid this, some algorithms perturb the
position and/or the velocity of the duplicated particles \cite{douc2005},
which without further adjustment results in the target moments not
being preserved. As the duplicated and perturbed particles evolve,
they will commonly diverge from one-another (assuming the particle
evolution is chaotic), and become independent samples of the
distribution after some time, with the divergence rate governing how
long this takes. Similarly, if particle evolution is stochastic, then
exactly duplicated particles will also diverge, and also become
independent samples of the distribution over time. In contrast, by
sampling from a binned approximation of $f$ and imposing a set of
integral constraints, MPCR up-sampling produces nearly independent
samples immediately.


For a second example, the resampling algorithm is applied to the
particle distribution taken from a tokamak plasma simulation
performed with the gyrokinetic PIC code XGC1.
In a spatially three-dimensional gyrokinetic PIC code, each particle
is characterized by six attributes, its weight $w$, spatial
coordinates $\mathbf{x} = (r;z;\phi)$, and velocity coordinate
$\mathbf{v} = (v_\|; v_{\perp})$. In this example, the distribution
$f$ is projected on to a spatially two-dimensional configuration space
in $r$ and $z$, taking advantage of statistical homogeneity in
$\phi$. The motivation for this aspect of the test, is to enable the
projection of a solution from a spatially three-dimensional plasma
turbulence simulation to serve as an initial condition for a
two-dimensional neoclassical simulation.  The data set consists of
$N^p=10^6$ particles, initially weighted to effect importance
sampling, as discussed in Section~\ref{sec:method}. However, particles
have since been mixed through their evolution, so their weights no
longer accomplish the desired importance sampling.
\begin{figure}[tp]
\centering
	\includegraphics[trim = 120mm 0mm 140mm 0mm, clip, width=.48\textwidth]{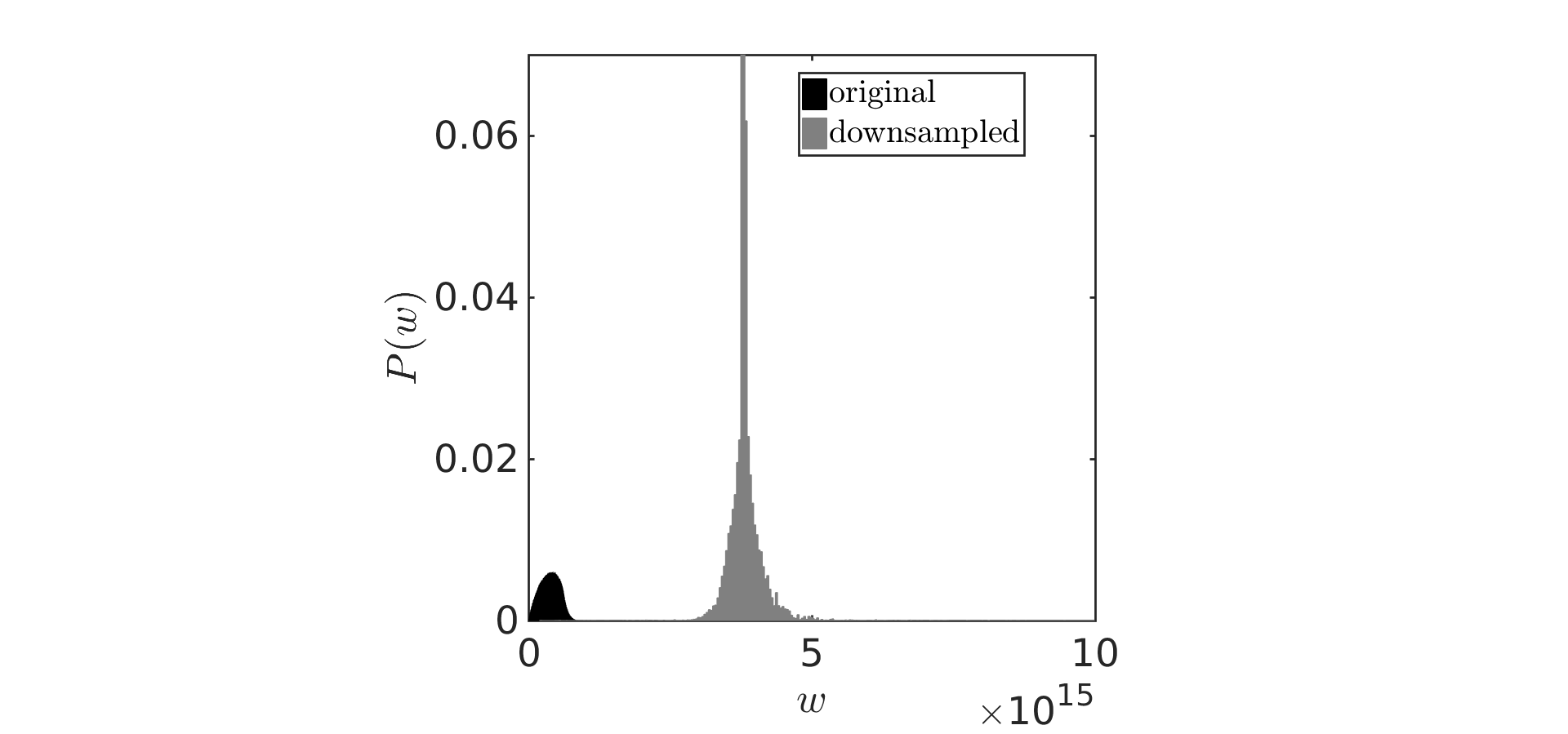}
	~
	\includegraphics[trim = 120mm 0mm 140mm 0mm, clip, width=.48\textwidth]{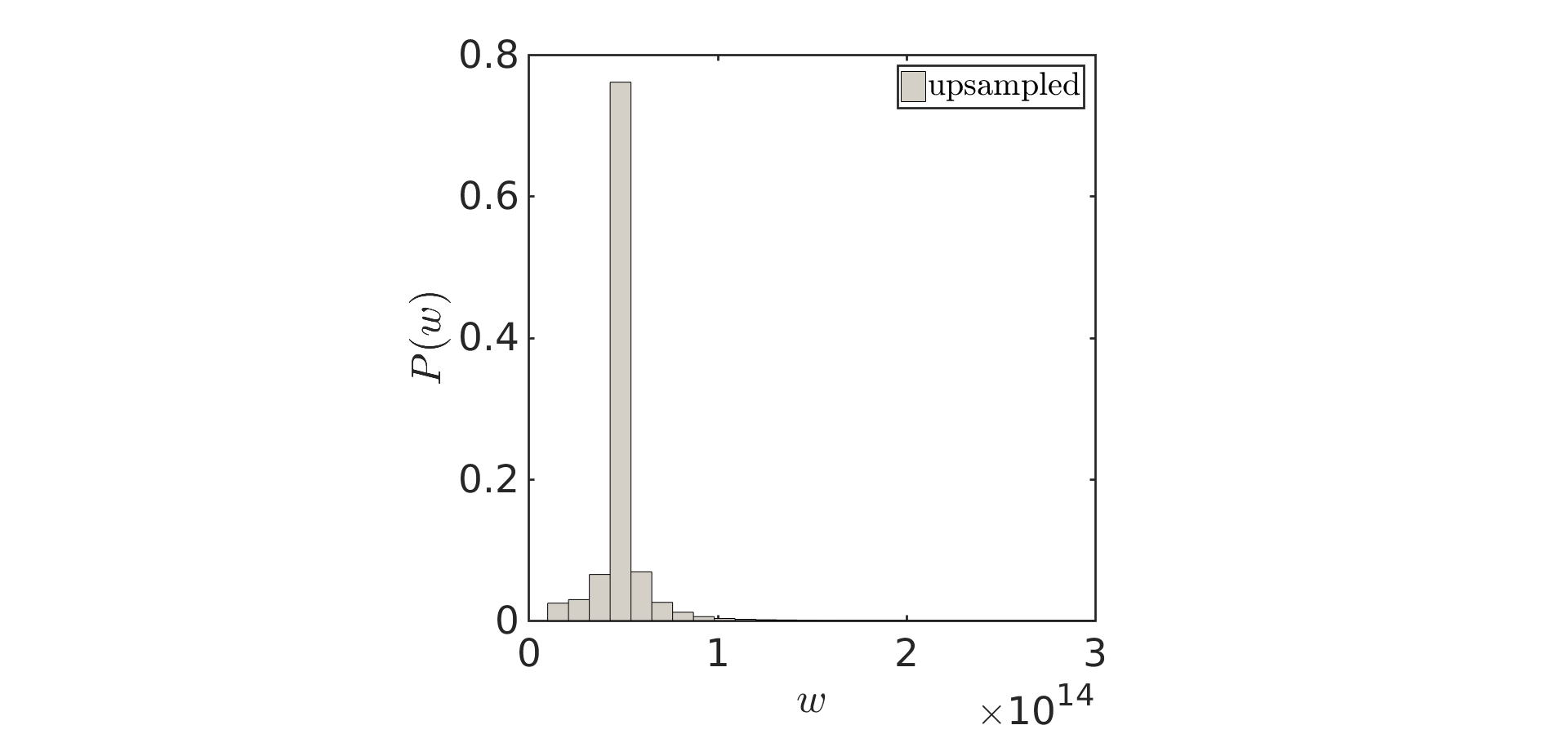}
\\  \hspace{0.0in} (a) \hspace{2.5in} (b) \vspace{-0.1in}
	\caption{
	Histograms of particle weights from the original
	weights and the proposed resampling algorithm. 
	In this case $N^p=10^6$ and the number of bins
	in the  $\rho$, $\mu$, $r$ and $z$ directions is 50:
	(a) original particles and downsampled particles with down-sampling factor (DSF) = 20,
	(b) upsampled particles with up-sampling factor (USF) = 10.}	
	\label{fig:hits_w}	
\end{figure}

The first step of the particle resampling process is to divide the
four-dimensional phase space $\boldsymbol{\zeta}$ into bins and sort
the particles into those bins.  In the gyrocentered XGC1 code, the
particle velocity space is parameterized in terms of the normalized
parallel velocity $\rho=v_\|/B$ and magnetic moment
$\mu=v_\perp^2/2B$, where $B$ is the magnitude of the magnetic
field. For this case then, the four-dimensional down-sampled phase
space is $\boldsymbol{\zeta}=(r,z,\rho,\mu)$. The $\boldsymbol{\zeta}$
domain is divided into bins, with 50 bins in each of the four
phase-space directions.

\blue{To demonstrate the adjustment of weights as part of the resampling
process the importance weighting is eliminated by making $g \propto
f$, where $g$ is the marker particle distribution function
in \eqref{eq:targetM} to enforce uniform target weights in bins.
}
The target number of particles $M^p_i$ in each
bin is then simply proportional to the sum of the weights of the
original particles in the bin. Specifically,
\begin{equation}
M^p_i=M^p\left(\frac{\sum_{j=1}^{N_i^p} w^i_j}{\sum_{j=1}^{N^p} w_j}\right).
\end{equation} 

One complication that arises in sorting the particles into bins is
that in some cases the target number of down-sampled particles $M^p_i$
in a bin is too small to allow the constraints to be imposed. In this
case, neighboring bins in the velocity space directions $(\rho,\mu)$
are merged to form larger bins with more particles, until $M^p_i$
exceeds a specified minimum, which was set here to $M^p_i \geq 25$.

The distribution of particle weights is shown in
Figure~\ref{fig:hits_w} for the original particles, and for the result
of down-, and up-sampling using the MPCR algorithm. Note that the
original distribution is broad, while the resampled distributions are
strongly peaked. The target marker particle distribution for both
cases was to eliminate importance sampling and make the weights
uniform. The fact that the resampled weight distributions are not
delta-functions is a consequence of the weight adjustments used to
preserve moments. The down-sampled particle weights are much larger
than the original weights and the up-sampled weights are much smaller
because the mean of the weights must increase (decrease) by the
down-scaling (up-scaling) factor.  Also by construction, the features
of the distribution in both space and velocity are preserved after
resampling. 

\begin{figure}[tp]
\begin{center}
\includegraphics[trim = 120mm 0mm 140mm 0mm, clip, width=.34\textwidth]{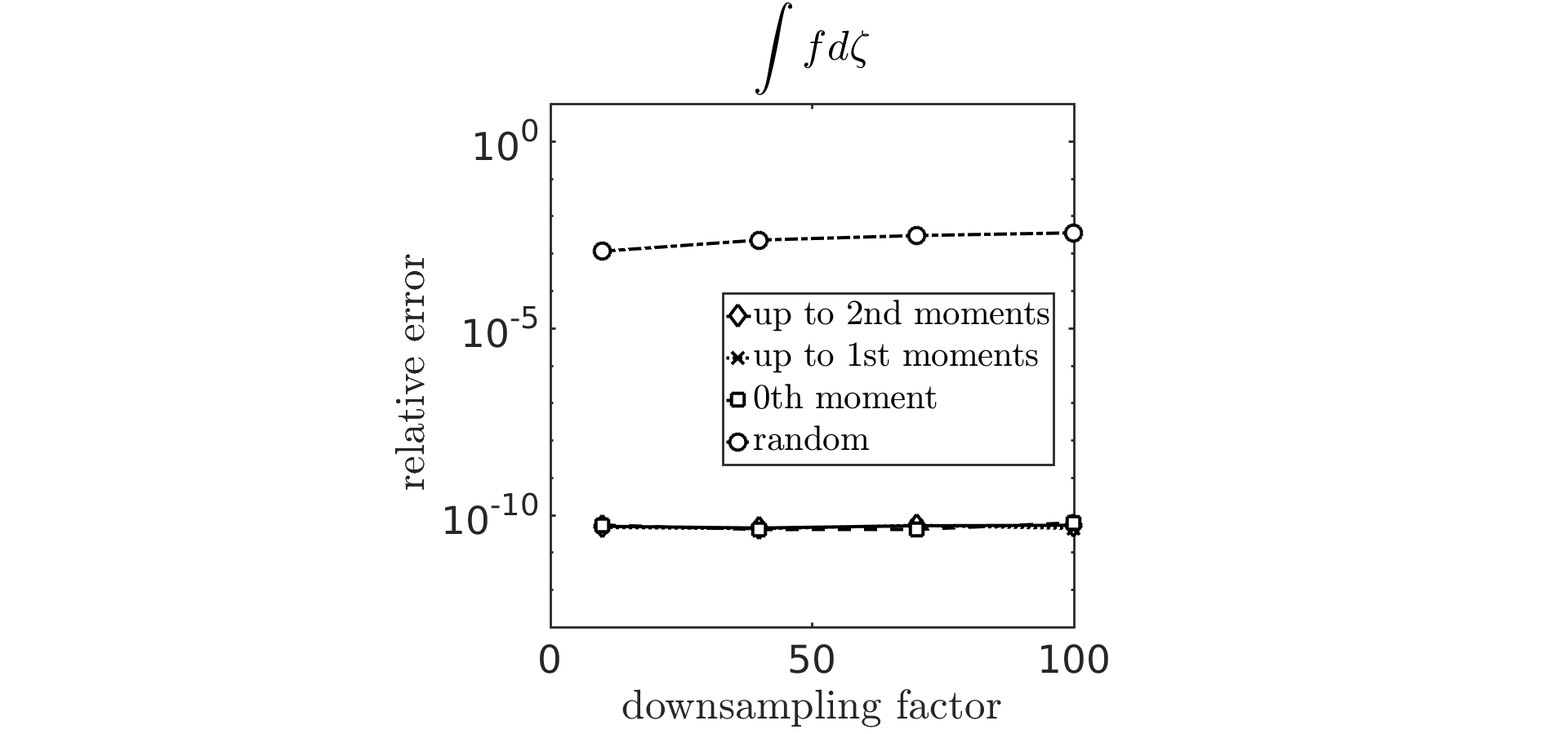}%
~
\includegraphics[trim = 150mm 0mm 140mm 0mm, clip, width=.31\textwidth]{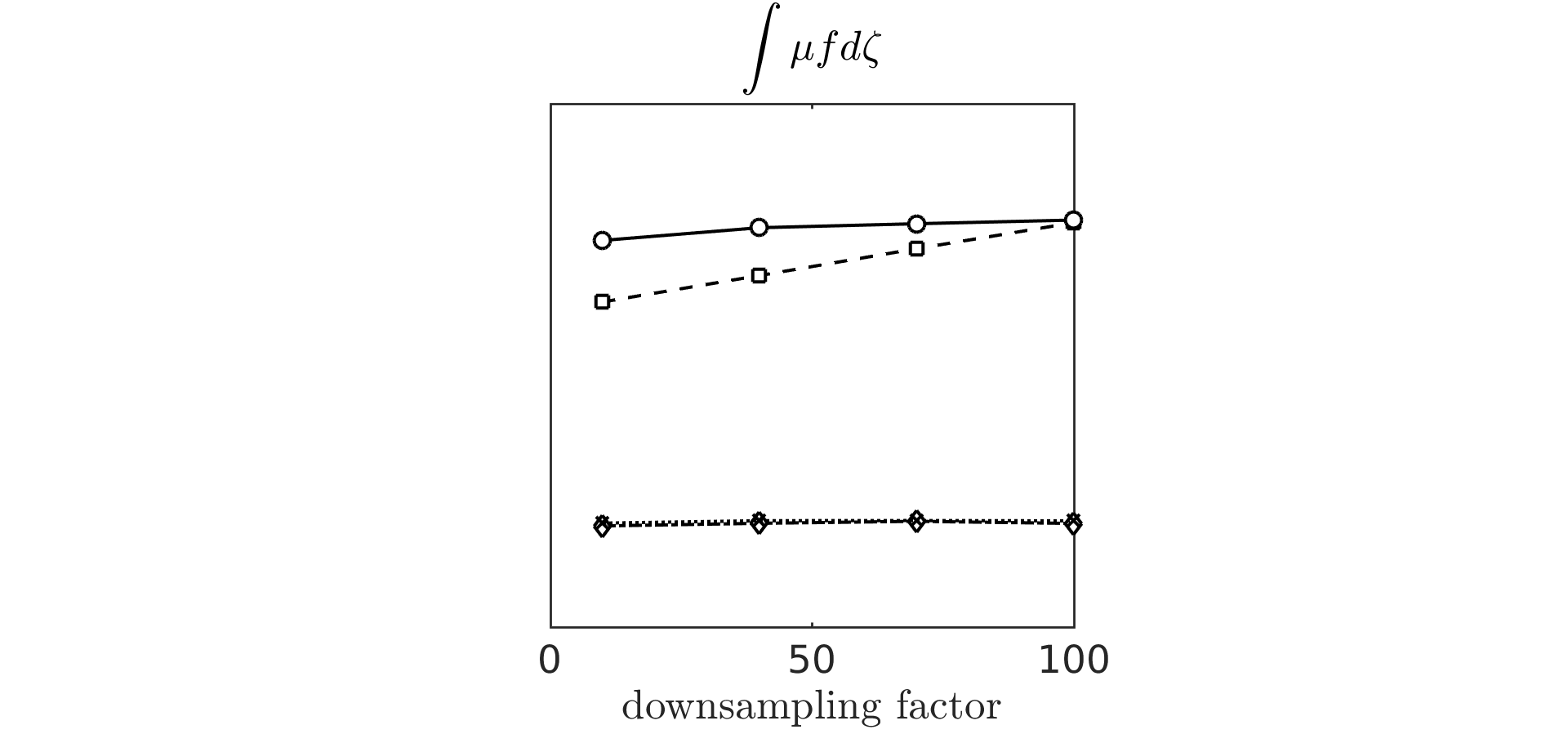}
~
\includegraphics[trim = 150mm 0mm 140mm 0mm, clip, width=.31\textwidth]{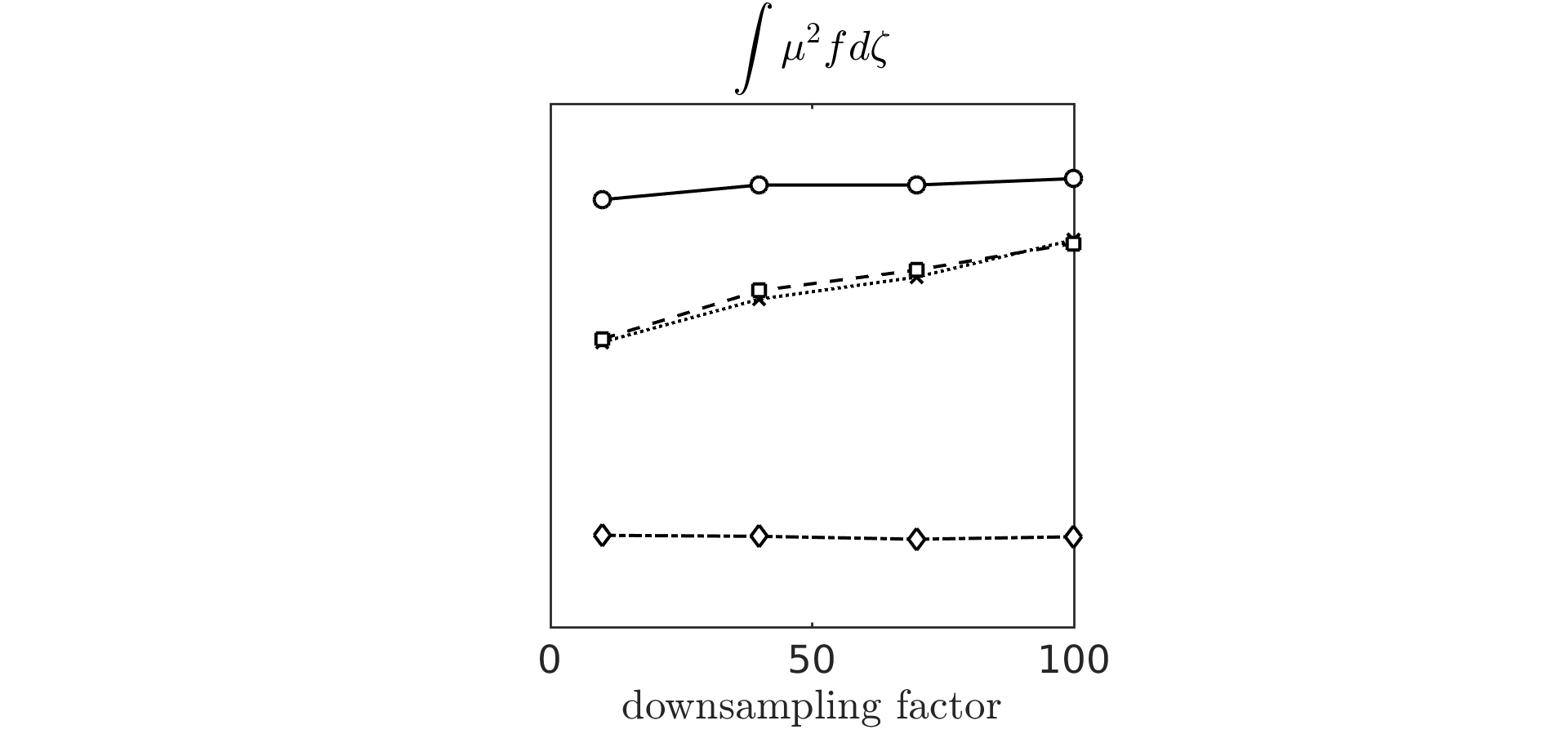}
\end{center}
\caption{Relative error in global zeroth, first and second velocity
	space moments as a function of down-sampling factor, using
	random down-sampling and MPCR down-sampling constrained to
	preserve zeroth, up to first and up to second moments.  Errors
	in configuration space moments are similar. There were
	$N^p=10^6$ original particles extracted from XGC1, and there
	were 10 bins in the $\rho$, $\mu$, $r$ and $z$ directions, for
	a total of $10^4$ bins.}  \label{fig:error_dsf}
\end{figure}

The accuracy of the moment preservation in the resampling approach is
tested by down-sampling while preserving different moments.  Three
cases were considered: preserving 1) the zeroth moment, 2) the zeroth
and first moments, and 3) the zeroth through second moments, as
described in Section~\ref{sec:method}. This was done for a range of
down-sampling factors, and resulting errors in the moments are shown in
Figures \ref{fig:error_dsf}, along
with errors for random down-sampling. The moment errors for the random
down-sampling are averaged over $10^4$ sampling realizations.

There are several interesting features of these errors. First, as
expected, the relative errors in the moments that are preserved are
order $10^{-10}$ or less. Second, with the MPCR algorithm, even the
moments that are not constrained have significantly lower errors than
for random sampling, in some cases by three orders of magnitude. This
improvement over random sampling is presumably due to the binning,
which enforces a discrete representation of the global distribution that
defines the moments. \blue{Notice in particular that the error in the second
moment when constraining the first moment is only marginally better
than when just the zeroth moment is constrained.}
Finally, for both random down sampling and for
unconstrained moments in MPCR, the relative error increases with
down-sampling factor. The observed weak growth is as expected for
random sampling, which should introduce errors that scale with the
square root of the down-sampling factor. In contrast the relative
errors in the unconstrained moments in MPCR grow much more rapidly
with down-sampling factor.

%

\begin{figure}[!t]
\begin{center}
	\includegraphics[trim = 0mm 0mm 0mm 0mm, clip, width=.32\textwidth]{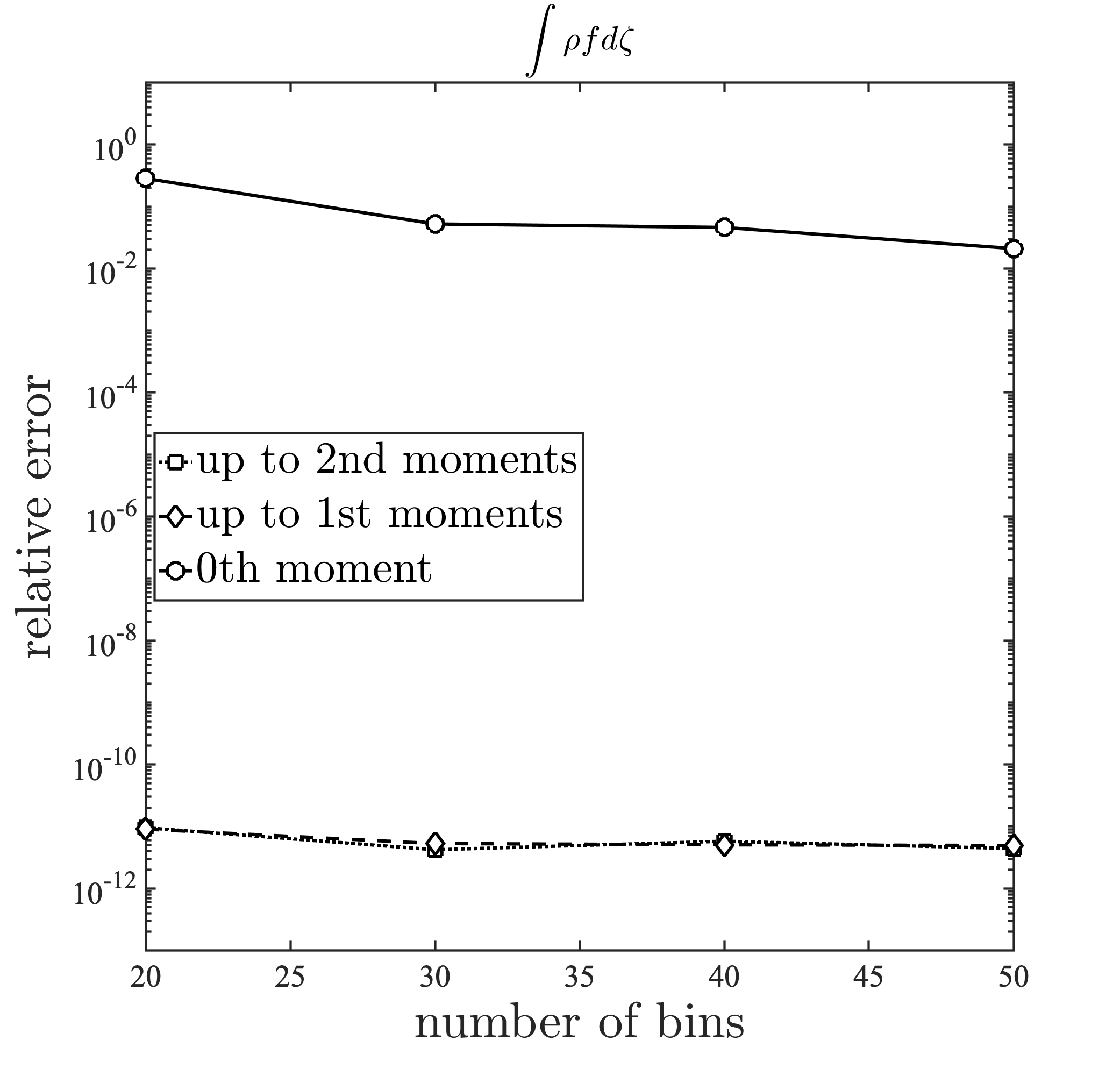}%
	\includegraphics[trim = 0mm 0mm 0mm 0mm, clip, width=.32\textwidth]{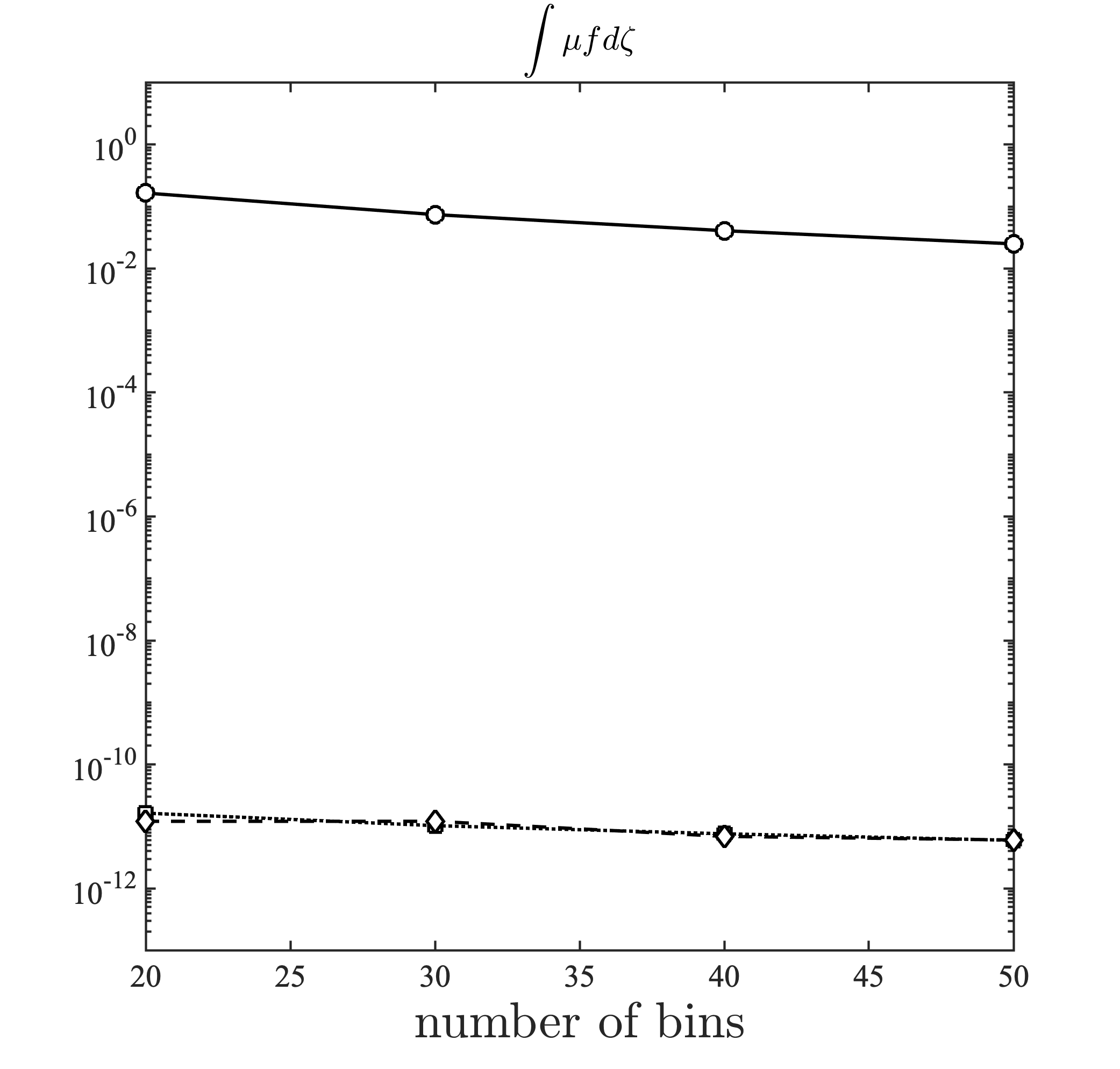}
	\includegraphics[trim = 0mm 0mm 0mm 0mm, clip, width=.31\textwidth]{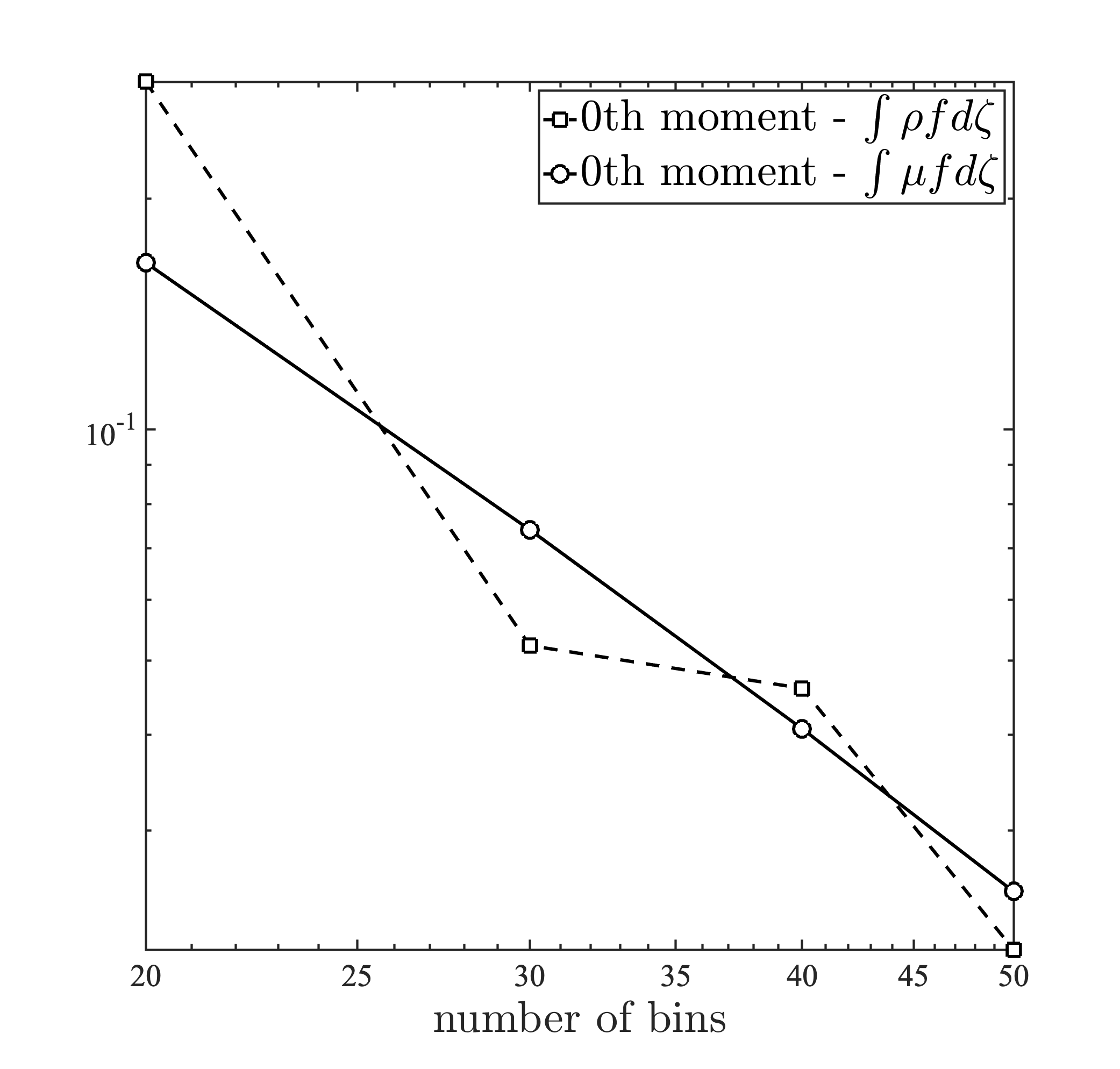}	
\end{center}
\caption{Relative
	error in global first moments in $\rho$ and $\mu$ as a
	function of the number of bins in each direction, $\rho$,
	$\mu$, $r$ and $z$.  MPCR up-sampling was used,
	constrained to preserve zeroth, up to first and up to second
	moments.  There were $N^p=10^6$ original particles extracted
	from XGC1, and the up-scaling factor was 10.  $E(\mu)\approx N^{-2.1}$ and $E(\rho)
	\approx N^{-2.7}$, where $N$ is the number of bins and $E(\cdot)$ is the error in the corresponding moment.	
	}
	\label{fig:UPerror_bin}	
\end{figure}

The reduction of errors relative to random down-sampling due to
binning, as discussed above, suggests that in MPCR, moment errors should
reduce with increasing bin density. This is indeed the case, as
shown in Figure~\ref{fig:UPerror_bin}, where relative error in an
up-sampling case is plotted as a function of the number of bins in
each direction. The error in the first moment when only the zeroth
moment is constrained appears to be decreasing like the number
of bins in each direction squared. Recall that the up-sampling
algorithm involves sampling a uniform distribution on each bin. As
mentioned in Section~\ref{sec:method}, more sophisticated sampling in
each bin can be used, which could result in a more rapid reduction of
errors with number of bins. Finally note that as with down-sampling,
preserved moments have relative errors that are better than $10^{-10}$.

\subsection{Periodic particle resampling of a neo-classical PIC simulation}

\blue{
In this subsection, we apply the MPCR algorithm to tokamak plasma
test cases using the XGCa code. For context, the code and simulation
characteristics are described briefly below, before presentation of
simulation results. XGCa \cite{hager2016_1,sku_2016,sku_2009} is a global gyrokinetic
particle-in-cell (PIC) code with an axisymmetric electrostatic
potential solver specialized for the simulation of neoclassical
transport physics in the edge plasma of toroidal magnetic confinement
devices.  These simulations are capable of evolving the full
five-dimensional gyrocenter distribution function from the magnetic
axis to the inner wall of the device using either conventional
full-$f$ (constant particle weight) \cite{sku_2009} or
total-$f$ (variable particle weights) \cite{sku_2016} methods.  
}

As a proof-of-principle test, the impact of the MPCR algorithm on
several important characteristics of the simulated plasma are
investigated here; particularly the radial electric field that is needed to
maintain quasi-neutrality, the maintenance of a divergence-free equilibrium
plasma flow, and the conservation of toroidal angular momentum.  For
simplicity, we use the full-$f$ representation of a single ion species (deuterium) with
the adiabatic electron model.

\blue{
The magnetic equilibrium field is that of a generic, up-down
symmetric, low-aspect ratio tokamak with a circular boundary surface
and Shafranov shift.  
Marker particles are initiated with a uniform distribution in
configuration and velocity space and perpendicular and parallel
velocities of up to approximately $3.5 v_{th}$, where $v_{th}$ is the
thermal velocity.  The marker particle weights $w$ are set to produce
a locally Maxwellian distribution with density $n_i(\psi)$ and
temperature $T_i(\psi)$ that depend only on the flux-label
(generalized minor radial coordinate) $\psi$; that is, each particle's
initial weight is a function of its initial flux-label and velocity.
The electrostatic potential is initially zero.
}

\begin{figure}[tp]
\centering
	\includegraphics[trim = 120mm 0mm 120mm 0mm, clip, width=.36\textwidth]{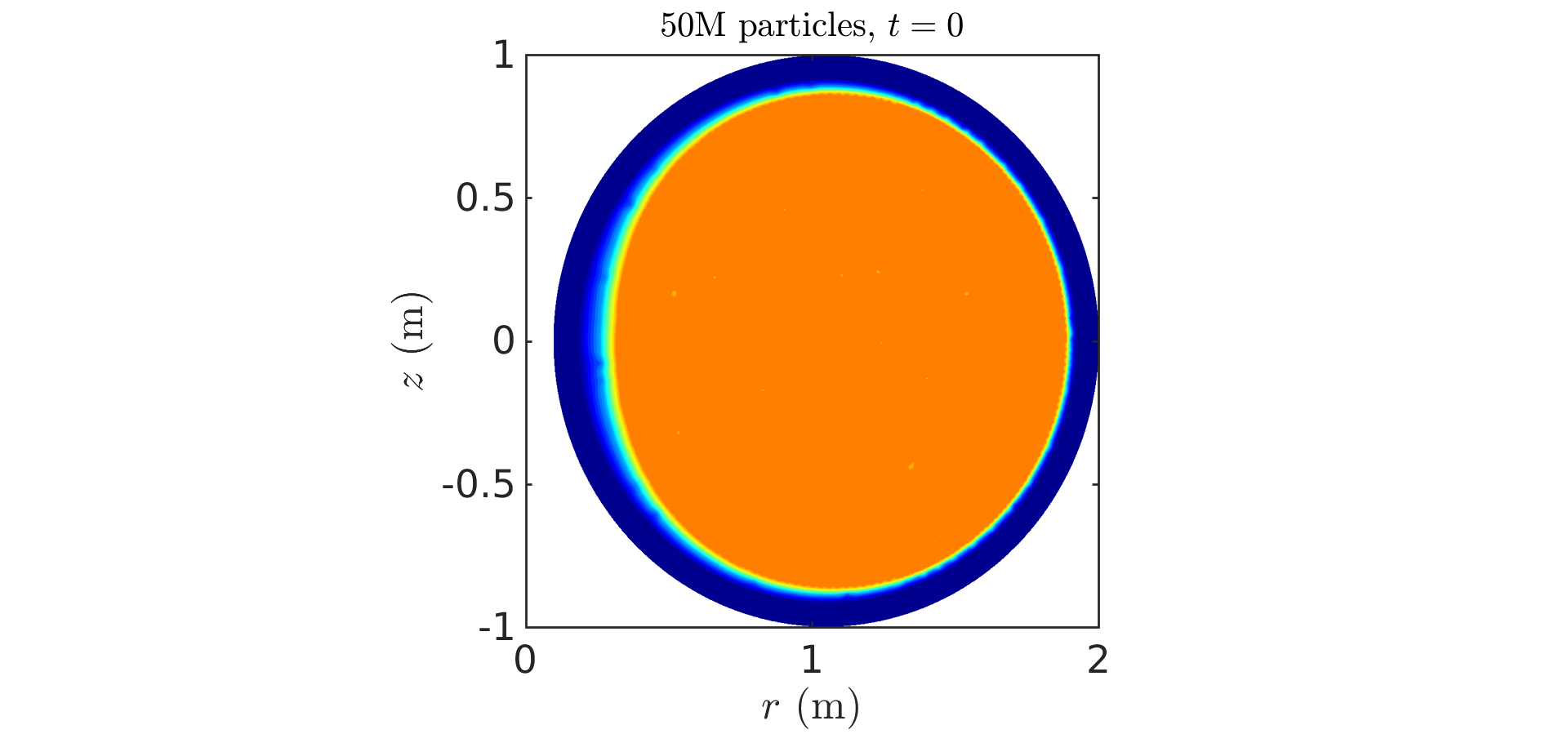}%
	\includegraphics[trim = 150mm 0mm 120mm 0mm, clip, width=.33\textwidth]{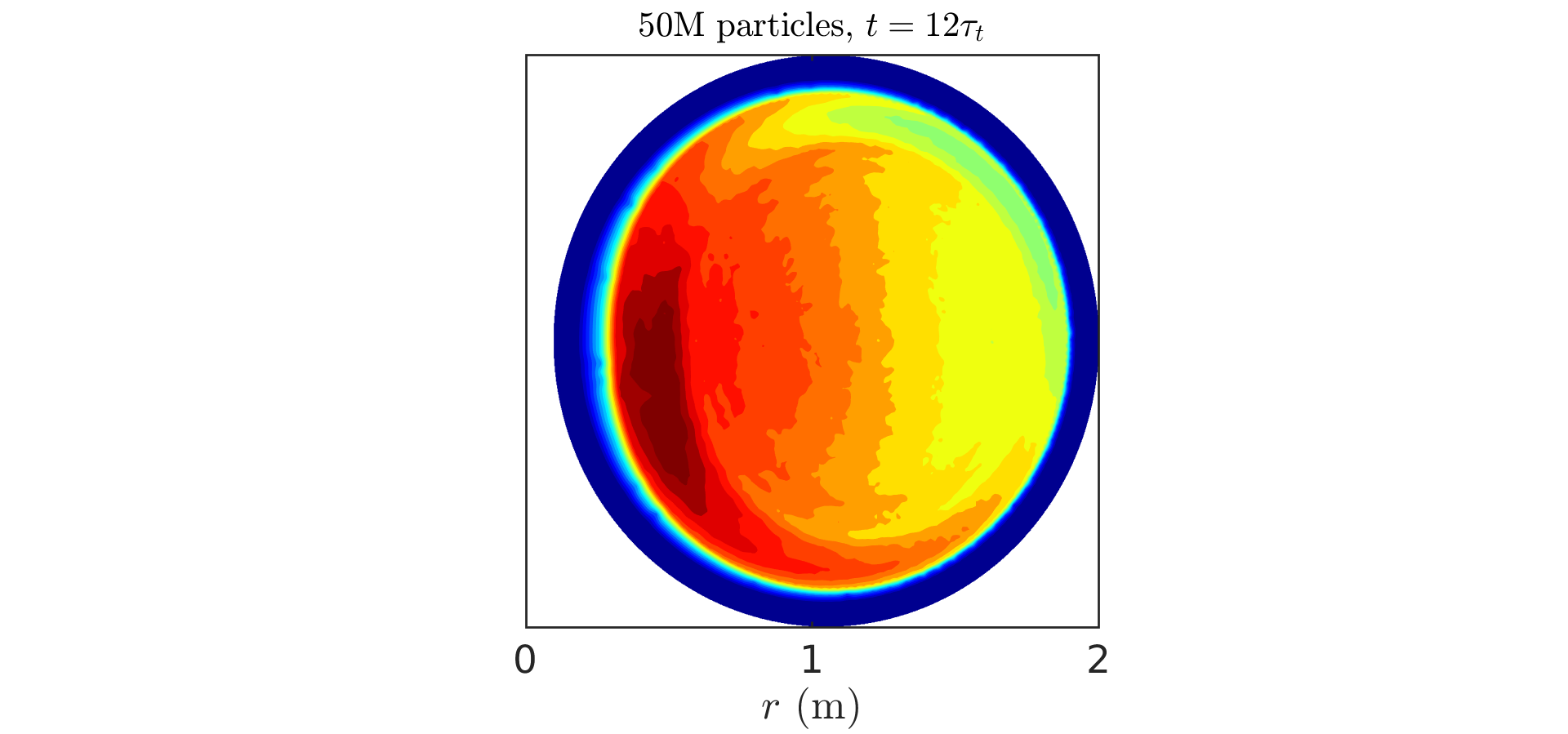}%
\includegraphics[trim = 150mm 0mm 100mm 0mm, clip, width=.345\textwidth]{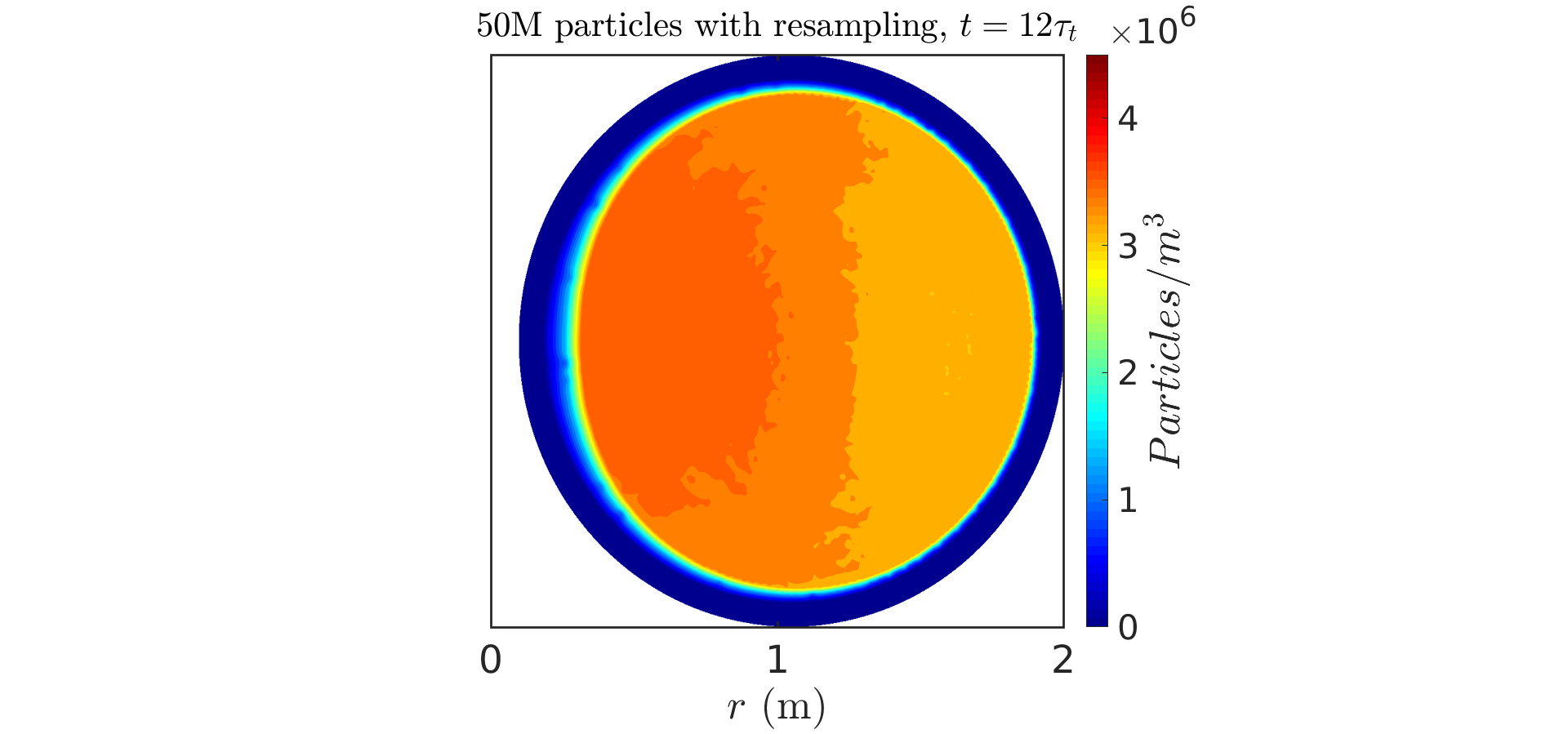}
	\caption{Spatial distribution of particle density for the
	simulation with 50 million particles: initial particle
	distribution (left), at $t=12\tau_t$ without resampling
	(middle), and at $t=12\tau_t$ with resampling (right).}	
	\label{fig:DensityDist}
\end{figure}

A local Maxwellian is, however, not the neoclassical equilibrium
distribution in the presence of a pressure gradient, due to the
magnetic inhomogeneity drift, which causes charged particles in
tokamaks to move on so-called banana-orbits with finite width.
Starting with a local Maxwellian distribution and
vanishing radial electric field, the orbital motion of the ions, together
with the background pressure gradient leads to the development of an
up-down anti-symmetric pressure variation and a net toroidal
flow. If this were the equilibrium flow it would
violate the divergence-free condition
and toroidal angular momentum conservation.  To make the
equilibrium flow divergence free and cancel the net toroidal flow from
the magnetic inhomogeneity drift, the plasma reacts by generating a
radial electric field (guiding center polarization) with its
corresponding $\mathbf{E}\times\mathbf{B}$ flow.

The challenge in calculating the radial electric field with a
full-f PIC code arises because the orbital motion of the markers
leads to mixing of particles with disparate weights.  This is a
considerable source of sampling error, especially in regions with
small pressure gradient and low amplitude of the electrostatic
potential.  The purpose of periodic particle-resampling is to reduce
the sampling error by locally homogenizing the particle weights \blue{and by
improving the distribution of marker particles in phase space.}  We
demonstrate this capability by comparing the time evolution of the
radial electric field and the radial electric field profile in
quasi-steady state between simulations with a total of 5 million, 50
million, and 500 million marker particles without particle
re-sampling, and a simulation with 50 million particles with periodic
resampling.
\begin{figure}[t]
\centering	
	\includegraphics[width=.33\textwidth]{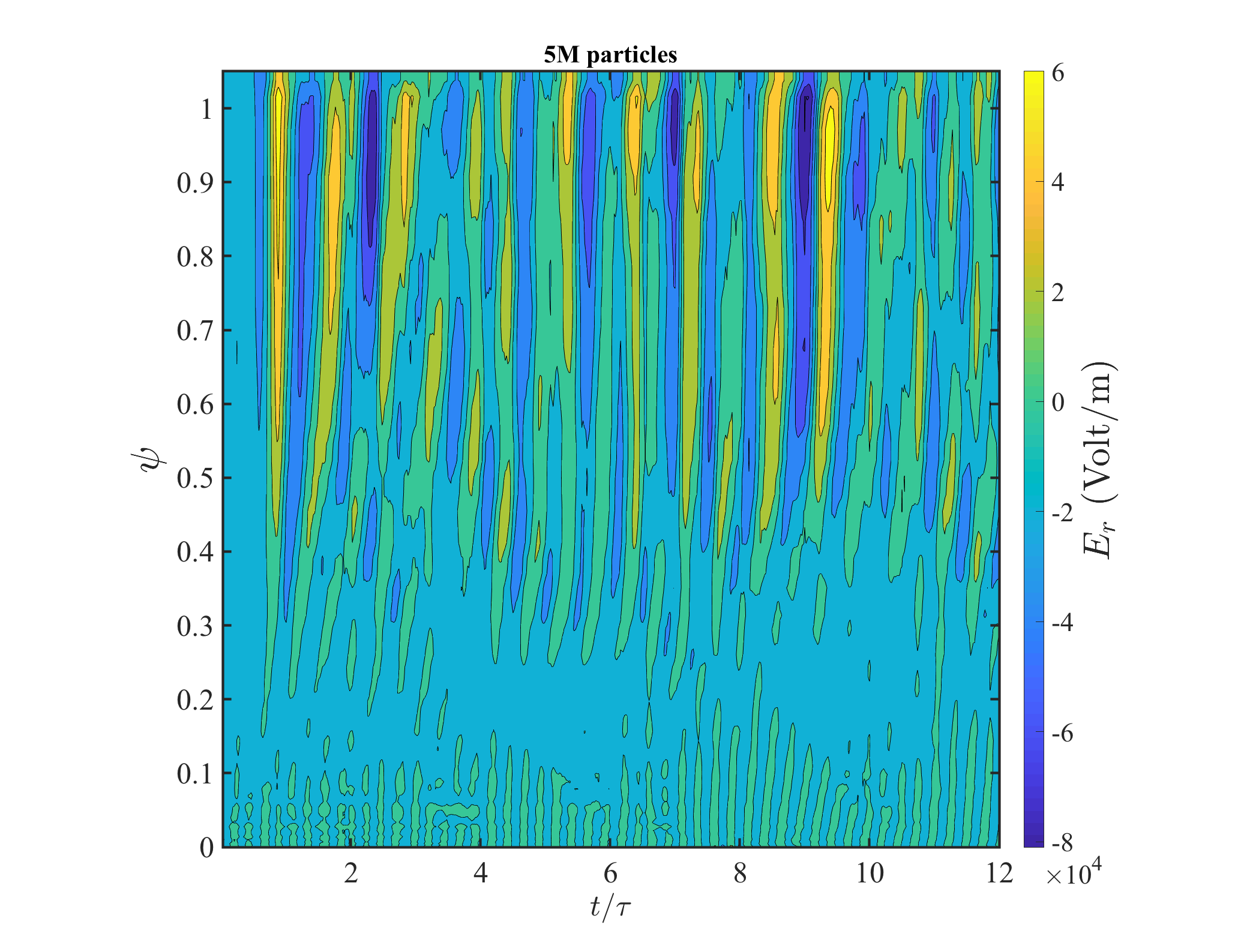}%
	\includegraphics[width=.33\textwidth]{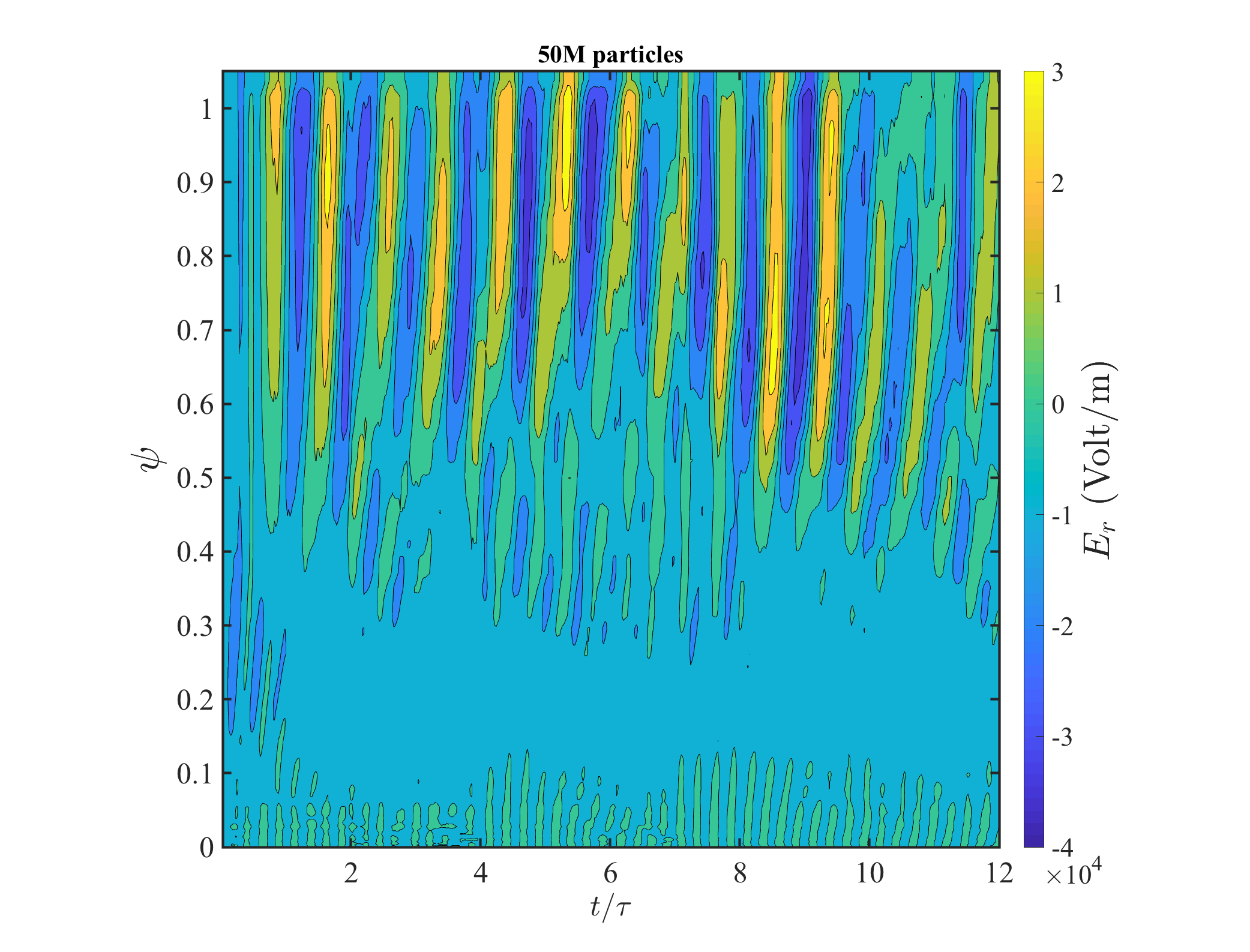}%
	\includegraphics[width=.33\textwidth]{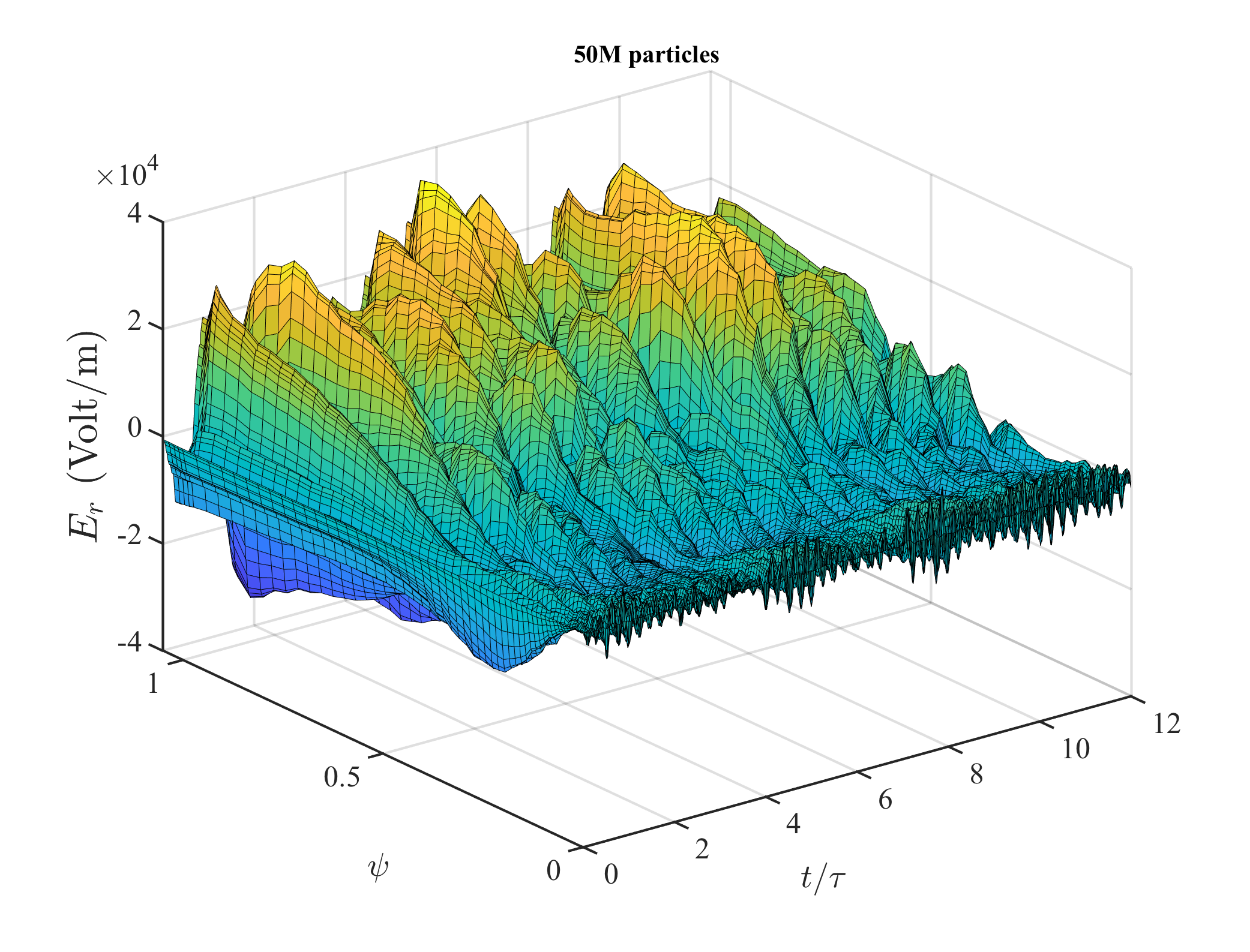}\\			
	\includegraphics[width=.33\textwidth]{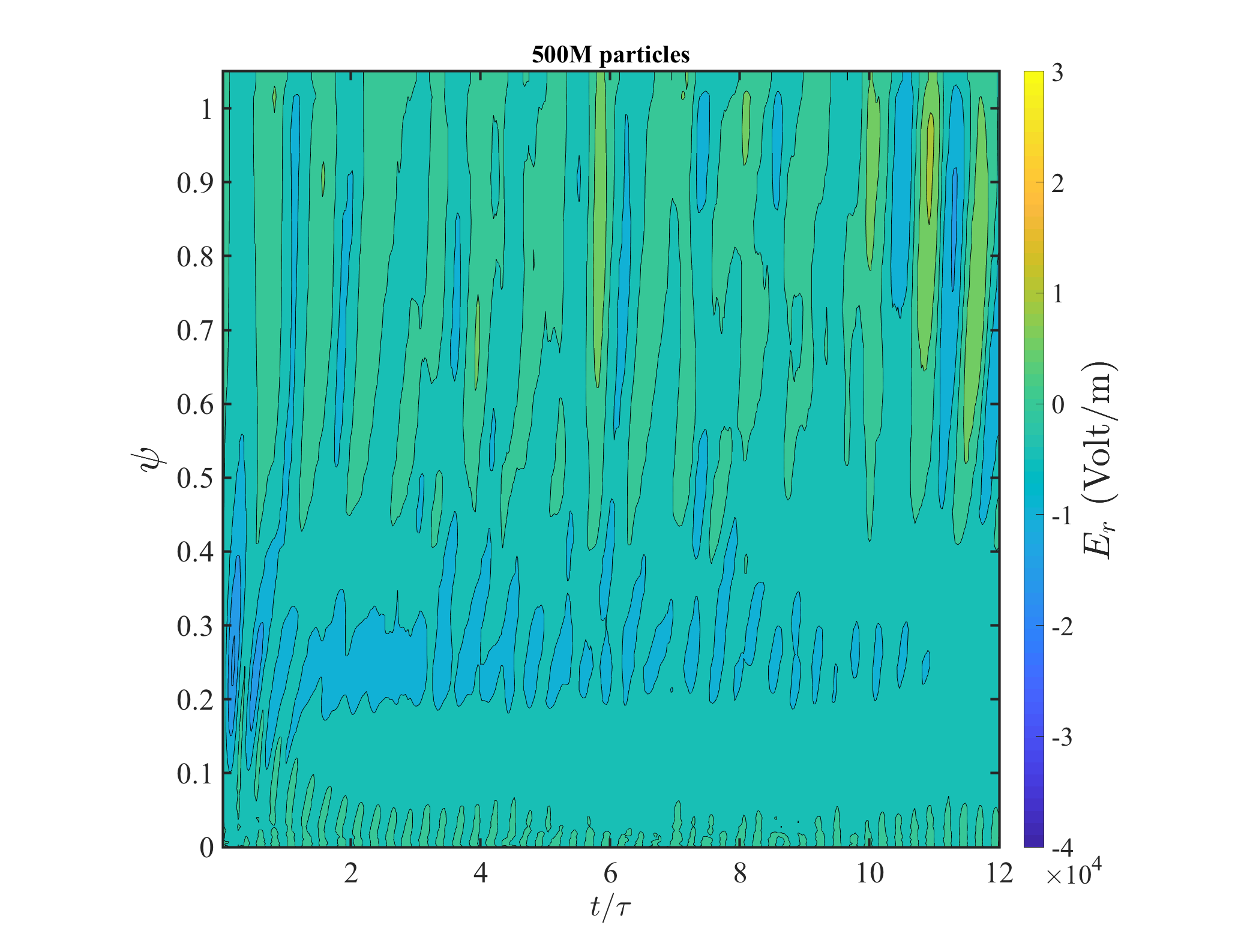}%
	\includegraphics[width=.33\textwidth]{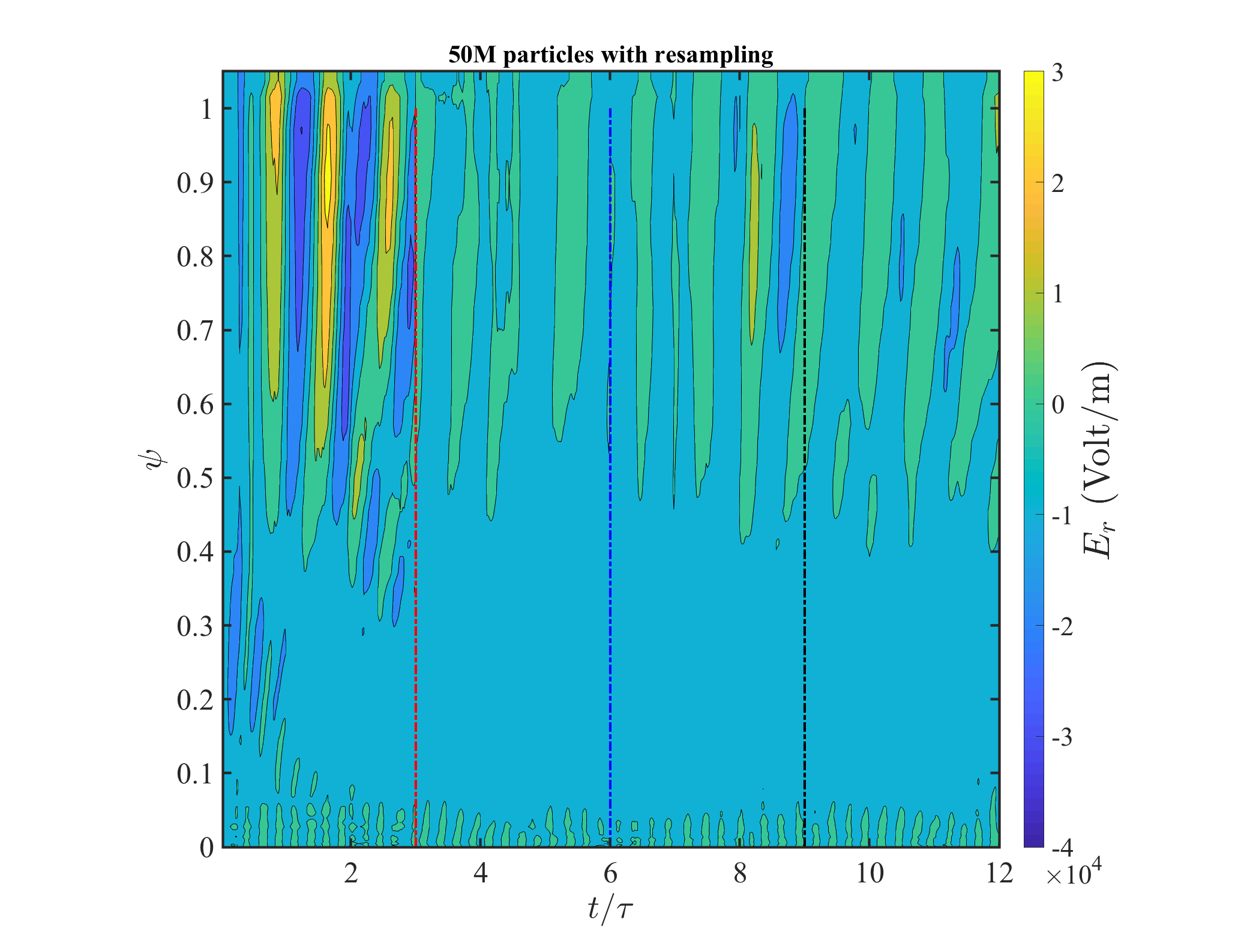}%
	\includegraphics[width=.33\textwidth]{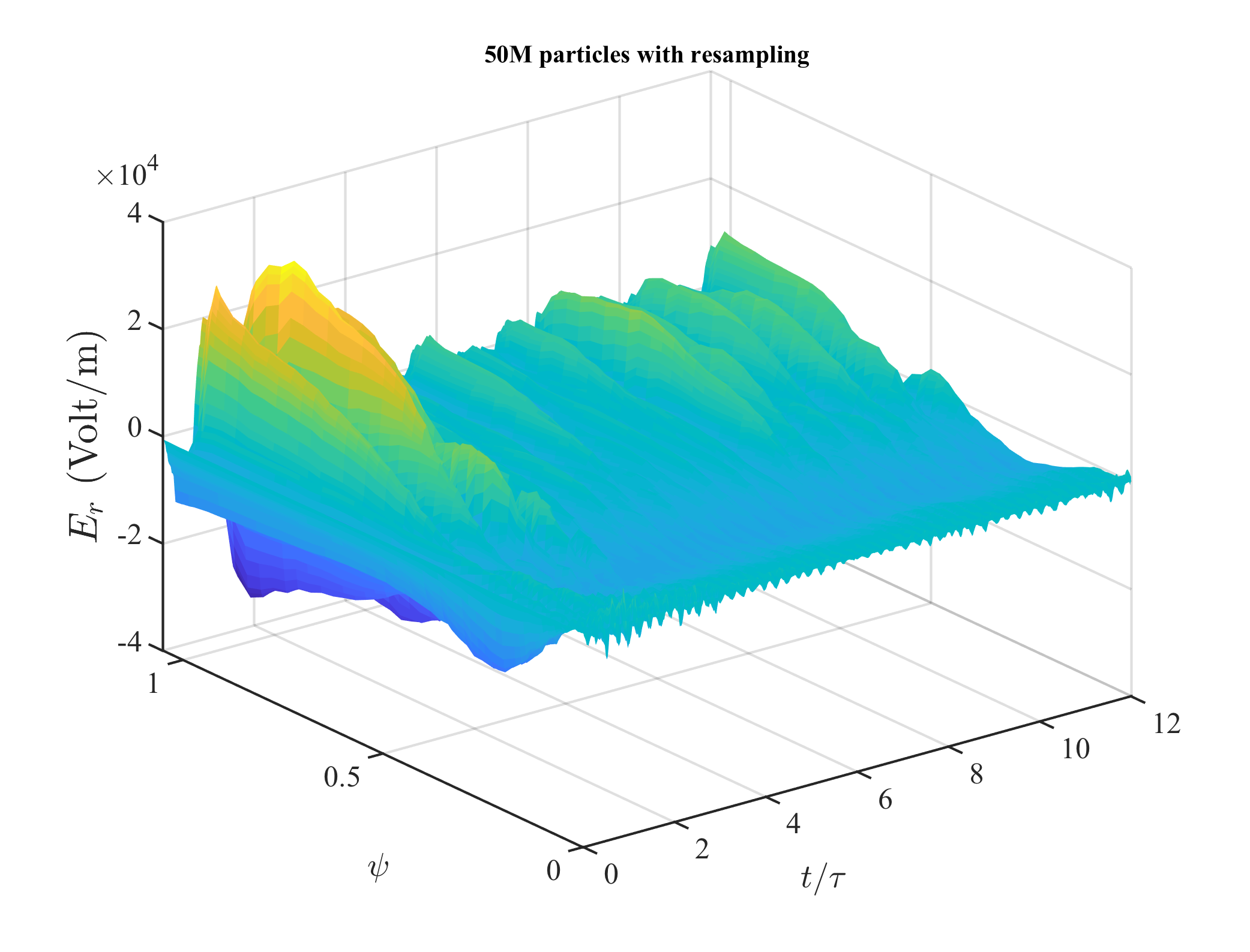}	
\caption{Spatio-temporal evolution of $E_r$ in simulations with 5, 50
	and 500 million particles without resampling and 50 million
	particles with resampling. 
        }	
	\label{fig:Er3d_5_50_500M}
\end{figure}

In all cases, the PIC simulations are run for 12 toroidal transit
times ($\tau_t$),
which is sufficient for the equilibrium plasma flow to
develop, \blue{and in this case the toroidally averaged equilibrium
solution is expected to be
slowly varying with no temporal oscillations.}
The spatial marker particle density is intended to be approximately
uniform through the core and edge plasma to yield a good
representation of the low-density edge region. However, as the plasma
evolves the particles mix and the initial marker distribution becomes
inhomogeneous, spoiling the intended importance sampling. This is
apparent in Figure~\ref{fig:DensityDist}, in which the initial and
final marker particle densities for the 50 million particle case are
plotted (no resampling). A similar mixing occurs in velocity space,
which would spoil any importance sampling used, for example, to represent high
energy ions. The result of
this poor sampling of the particle distribution is that large Monte Carlo
sampling error is introduced into the simulation. As an example,
consider the spatio-temporal evolution of the radial electric field
(Figure \ref{fig:Er3d_5_50_500M}) for the 5, 50 and 500 million
particle cases (no resampling).  In all three cases, \blue{there are spurious
oscillations, which} are
largest in the edge region, $0.8< \psi <1$, because the marker
particle densities are low in this region. Further, \blue{the magnitude of
these oscillations} is reduced with increasing number of 
particles. Particularly, the oscillation magnitude decreased by about a
factor of $\sqrt{10}$ going from 50M and 500M particles, which is the
expected asymptotic convergence rate for errors dominated by sampling
error.


To reduce the sampling error, the MPCR algorithm was applied
periodically in the 50 million particle simulation, in this case every
3 toroidal transit times ($3\tau_t$). There are 1500 times steps
between resamples, and the computational cost of resampling is
negligible compared to the time advance for this number of steps.  As
described in Section~\ref{sec:XGCimp}, the 2-dimensional unstructured mesh in
XGC is used for the configurational space bins consists of 6368 elements, and in velocity space 100
uniform bins were used in both the $v_\parallel$ and $v_\perp$
directions. At each resampling, the total number of marker particles
remained the same ($M^p=N^p$), and the target number of particles in
each bin is consistent with the initial marker particle distribution
(at $t=0$).

\begin{figure}[tp]
\centering
	\includegraphics[width=.5\textwidth]{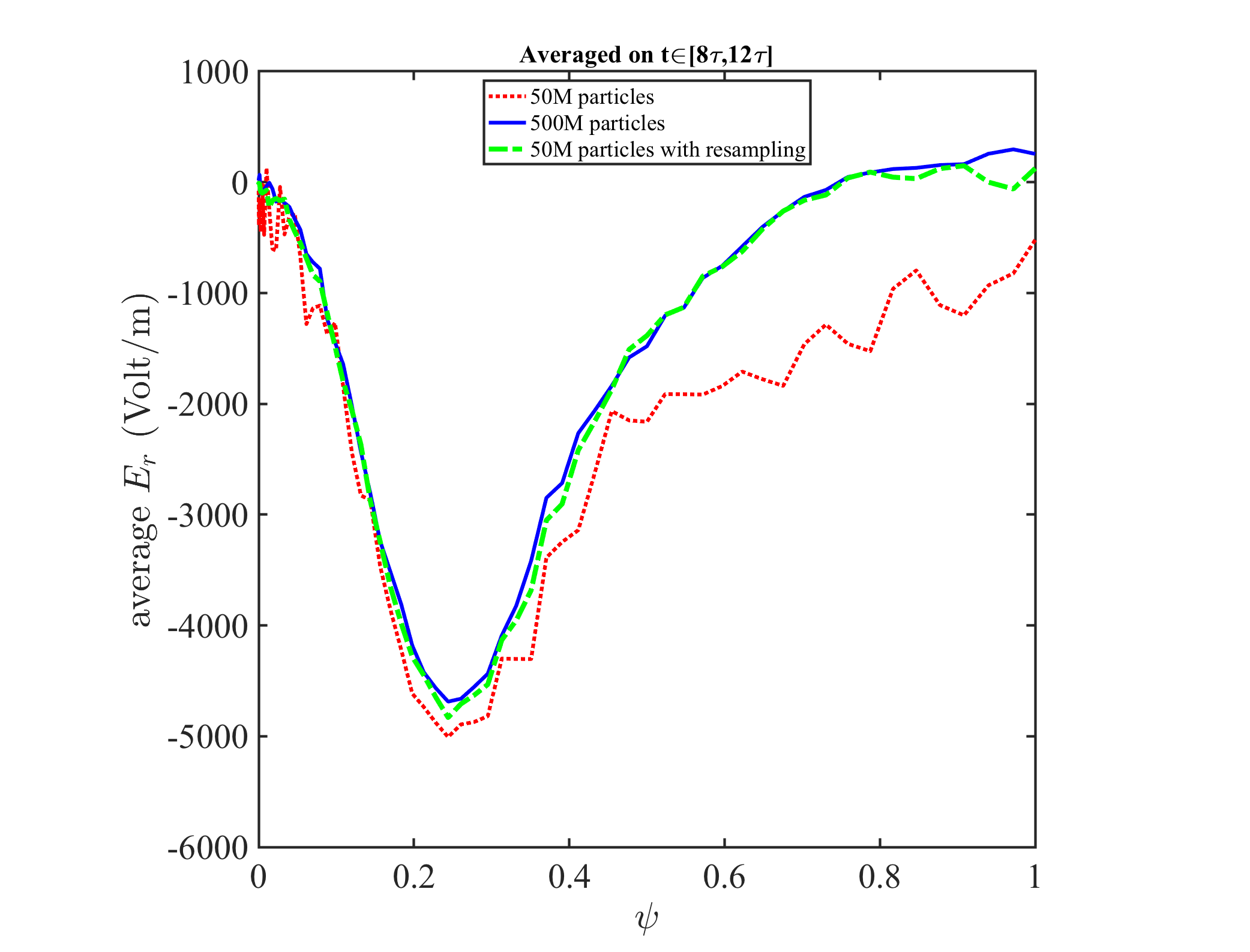}%
	\includegraphics[width=.5\textwidth]{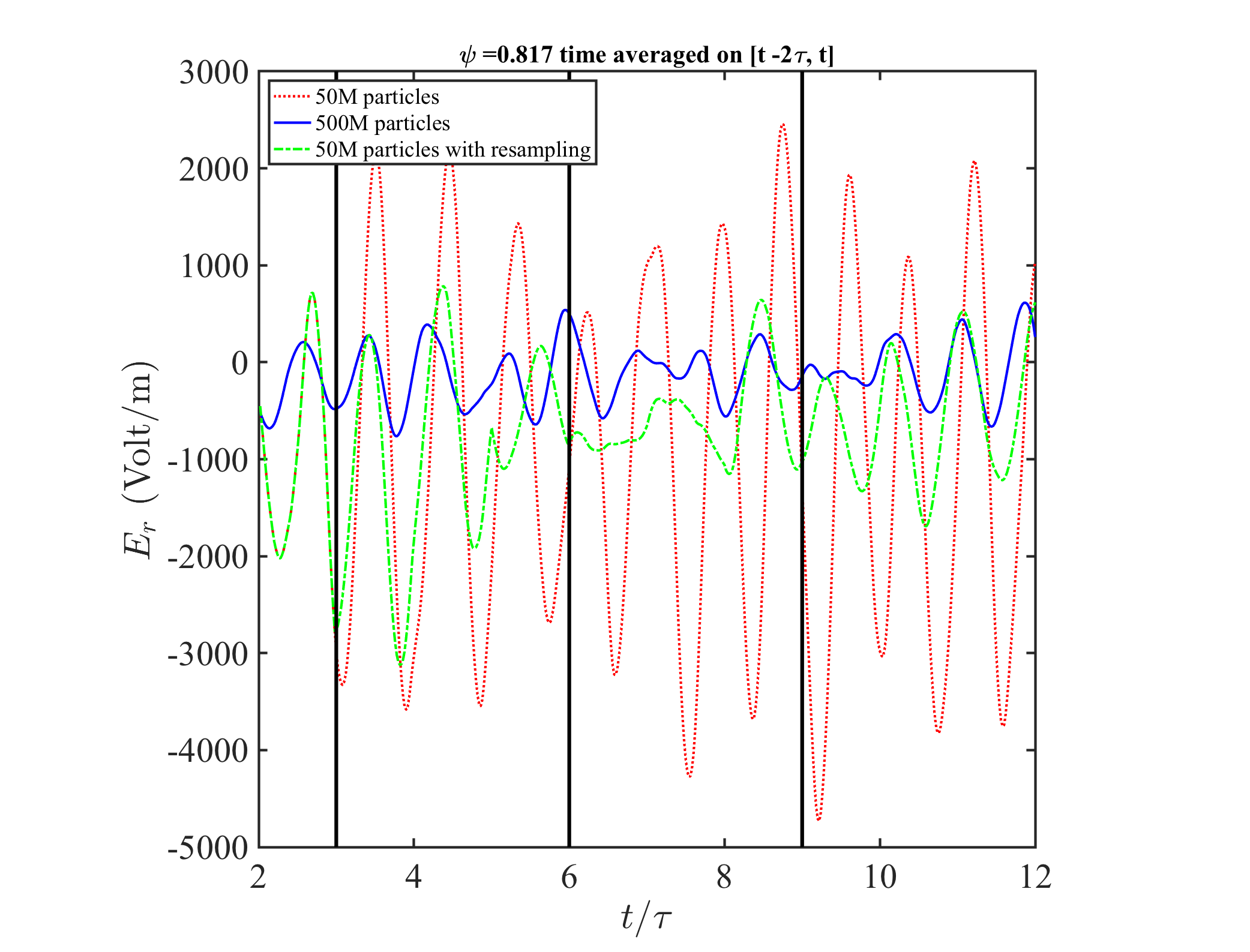}
	\\
	\includegraphics[width=.5\textwidth]{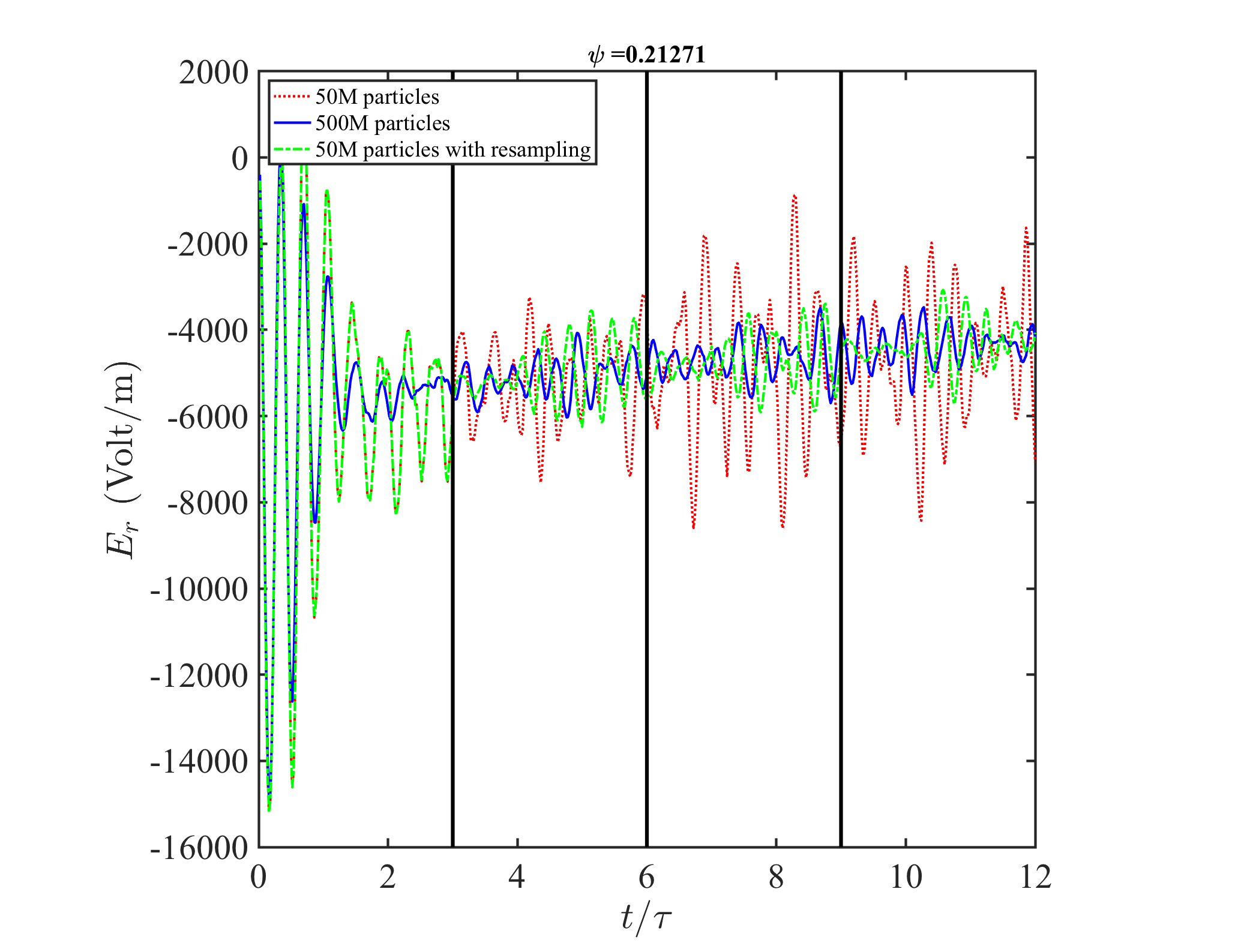}%
	\includegraphics[width=.5\textwidth]{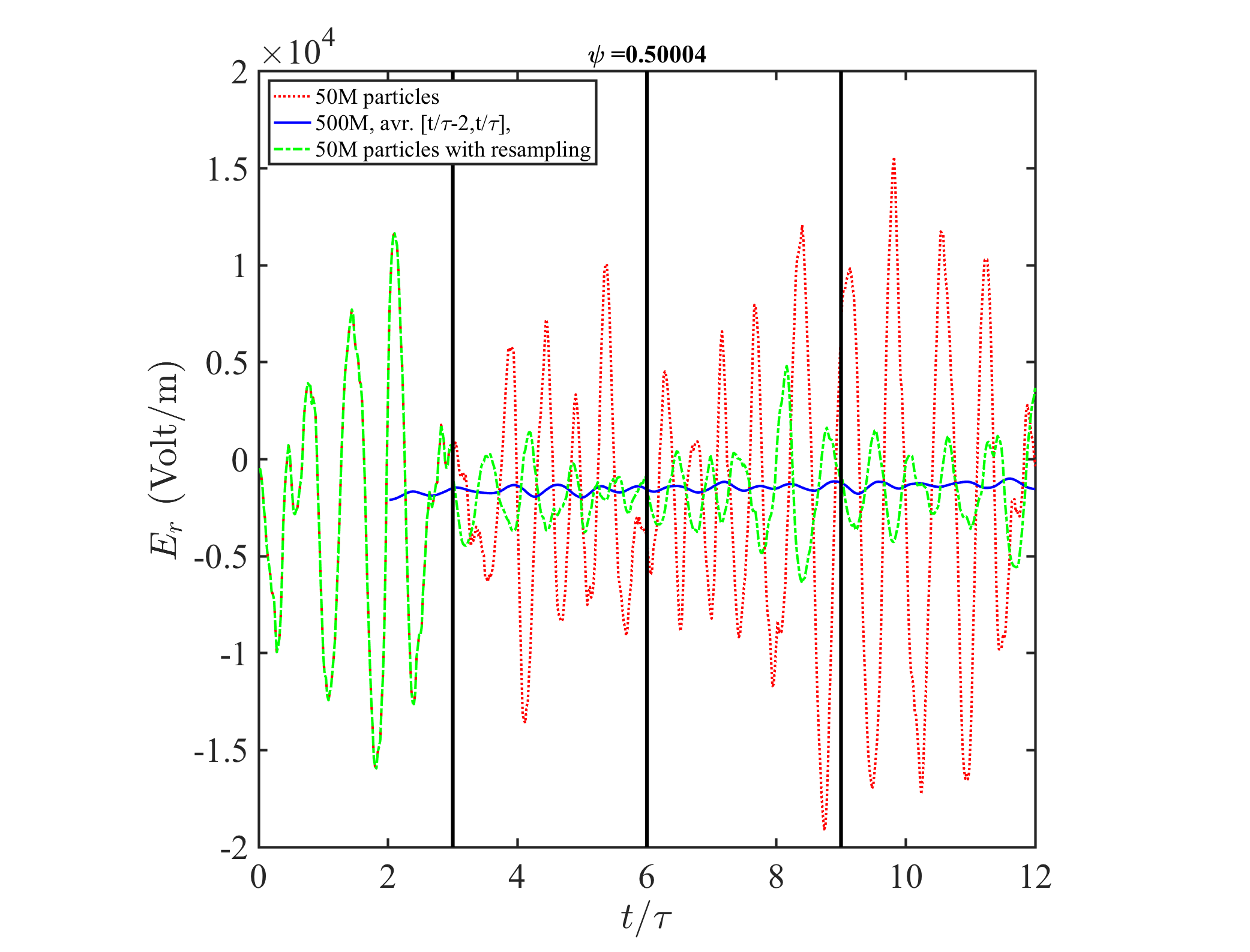}
	\caption{ $E_r$ profile at $t=12\tau_t$ and Time evolution of
	$E_r$ at $\psi = 0.5$, $\psi = 0.2$, $\psi = 0.8$, for
	simulations with 50 million particles with and without resampling,
	and 50 million particles. Resampling occurs at the times marked with
	a vertical black line.} \label{fig:Er1d}
\end{figure}
As a result of the periodic resampling, the marker particle
distribution in configuration space at the end of the simulation is
much more uniform than without the resampling
(Figure~\ref{fig:DensityDist}). This leads to much smaller oscillations in the
radial electric field $E_r$ (see
Figure~\ref{fig:Er3d_5_50_500M}). The resampled and non-resampled cases
are identical until the first resampling at $t=3\tau_t$. After the
resampling the spurious oscillations in $E_r$ are immediately reduced.  Indeed, it is
reduced to levels comparable to the 500M particle non-resampled case,
as can be seen in Figure~\ref{fig:Er1d} where the time evolution of
$E_r$ at several points in $\psi$ is plotted for the both 50 million
particles cases (resampled and not) and the 500 million particle
case. Also shown is the $E_r$ profile as a function of $\psi$ averaged
over the period $8\tau_t<t<12\tau_t$, and the profile from the 50
million particle resampled case is a much more accurate approximation
of the 500 million particle profile than the non-resampled case.
\blue{As an indicator of the oscillations in the PIC simulations 
Table \ref{tab:stDev_Er} shows the time averaged standard deviation  
of the $E_r$ for different time range.}

After each resampling, the spurious oscillations grow slowly as can be
seen between the resamplings at $t=6\tau_t$ and $9\tau_t$. However, in
the initial period, the \blue{oscillation amplitude becomes large
almost immediately} so that by $t\approx 3\tau_t$, they reach the
magnitude maintained throughout the non-resampled simulation. This is
presumably due to the relatively rapid evolution of the system toward
the equilibrium plasma flow, which would rapidly mix the particles in
both configuration and velocity space. Apparently once the flow is
near equilibrium, the mixing rates are more modest resulting in slower
growth of the sampling errors and resulting oscillations. This suggests
that rather than resampling at a fixed frequency, it would be useful
to resample locally and adaptively based on local estimates of the
sampling errors. This would be particularly useful in simulating
transient phenomena, and could adaptively determine the required
number of particles to attain a specified sampling error tolerance.

In PIC simulations, the primary computational cost is the evolution of
particles, and the relative cost of periodic particle resampling in a
plasma PIC simulation will be small.  
\blue{
The primary expense in resampling is the setup and solution of the
quadratic program. The dominant cost there is the QR factorization of
the constraint matrix.  This cost scales quadratically with the number
of constraints. For the XGCa test case given above, this results in a
computational complexity roughly 20 times that of a XGCa timestep
(using an explicit RK4 time integrator).  Given that the resampling
occurs infrequently, in this case every 1500 steps, the resampling
overhead is negligible (roughly 1\% of the simulation cost).  This
computational complexity estimate over-estimates the relative cost of
the resampling because it does not account for the cost of
interprocess communication, which is important for the XGCa time
advance but not resampling, because the resampling algorithm is
completely local.  Accurate timing data to more precisely characterize
the cost impact of resampling in the context of XGCa requires full
integration of the resampling algorithm into the XGCa software, which
is currently being developed.}
%
The
results presented here indicate that application of MPCR will allow
significantly fewer particles to be used to attain the same accuracy,
thereby reducing the cost of PIC simulations, which scale
approximately with the number of particles.

\begin{table}[]
\caption{\blue{The time average standard deviation of $E_r$ in Figure \ref{fig:Er1d}.}}
\label{tab:stDev_Er}
\begin{tabular}{|c|ccc|ccc|ccc|}
\hline
\multirow{2}{*}{Time range}                   & \multicolumn{3}{c|}{$\psi = 0.8$}     & \multicolumn{3}{c|}{$\psi = 0.5$}     & \multicolumn{3}{c|}{$\psi = 0.5$}      \\ \cline{2-10} 
                                              & 500M$^*$ & 50M$^{**}$   & 50M+res$^{***}$ & 500M  & 50M  & 50M+res & 500M   & 50M  & 50M+res \\ \hline
$3 < t/\tau < 6$  & 2980 & 16400 & 2350                & 673   & 2600 & 699                 & 380.61 & 1450 & 442.3               \\
$6 < t/\tau < 9$  & 2660 & 17900 & 4690                & 711   & 2950 & 769                 & 210.87 & 2210 & 342.71              \\
$9 < t/\tau < 12$ & 4682 & 15100 & 5880                & 464.5 & 2230 & 714                 & 400    & 1540 & 452.32              \\ \hline
\end{tabular}
$^*$500 million particles.\\
$^{**}$50 million particles.\\
$^{***}$50 million particles with periodic resampling.
\end{table}

%% file: paper_5_conclusions.tex
The MPCR (Moment Preserving Constrained Resampling) algorithm
presented here is designed to reduce the sampling error introduced by
particle representations of distribution functions in PIC
simulations, while preserving important features of the
distribution. The algorithm functions by first partitioning the
particle phase space into bins. It then uses sampling techniques
within each bin to generate new particle positions and velocities, and
employs constrained optimization to adjust the particle weights in
each bin to preserve essential moments of the distribution function.
These can include such physically relevant quantities as mass,
momentum, energy, current density and charge density. Further, for
quantities that are projected onto the Eulerian grid in the PIC
method, the preserved quantities include the projections onto each
Eulerian degree of freedom. In this way, the Eulerian solution is
unchanged by the resampling process. Finally, by performing the
resampling in bins rather than globally, MPCR can be used to maintain
the desired marker particle distribution and importance sampling
weights, despite mixing and particle weight evolution due to the use of
variance reduction sampling.



The quality and utility of the MPCR algorithm was demonstrated here
through several examples. In application to a synthetic distribution
and to particles extracted from the gyrokinetic plasma PIC code XGC1,
MPCR was shown to preserve extraordinary features of the distribution
and to preserve various moments to the level or roundoff error. Using
a neoclassical test problem, it was also shown that by periodic
resampling using MPCR, the particle sampling error can be
significantly reduced, to a level comparable to a simulation with an
order of magnitude more particles.
\blue{Further, the computational cost of the periodic resampling is
negligible compared to the cost of the PIC simulation.}  Therefore,
these tests indicate that the use of MPCR resampling in a PIC code can
reduce computational cost and increase solution accuracy.  While not
pursued here, it was also noted that MPCR resampling can be applied
locally and adaptively to ensure that sampling errors remain below
specified tolerances.
